\documentclass[paper]{JHEP3} 

\JHEPspecialurl{http://jhep.sissa.it/JOURNAL/JHEP3.tar.gz}

\usepackage{epsfig,multicol}
\usepackage{cite}

\usepackage{amsmath}
\usepackage{amssymb}
\preprint{
KUNS-1913\\
KEK-TH-957\\
hep-th/0405277\\}

\title{Dynamical aspects of the fuzzy CP$^{2}$ 
in the large $N$ reduced model with a cubic term}
\author{ Takehiro Azuma${}^a$, Subrata Bal${}^{b,c}$, 
Keiichi Nagao${}^{a,d}$, Jun Nishimura${}^{a,e}$ \\
\llap{$^a$}
High Energy Accelerator Research Organization (KEK),\\
1-1 Oho, Tsukuba 305-0801, Japan  \\
\llap{$^b$}Department of Physics, Kyoto University, 
Kitashirakawa,\\
Kyoto 606-8502, Japan\\
\llap{$^c$}School of Theoretical Physics,
Dublin Institute for Advanced Studies,\\
10 Burlington Road, Dublin 4, Ireland \\
\llap{$^d$}Theoretical Physics Laboratory, College of Education, \\
Ibaraki University, 2-1-1 Bunkyo, Mito, Ibaraki 310-8512, Japan \\
\llap{$^e$}Department of Particle and Nuclear Physics,\\
Graduate University for Advanced Studies (SOKENDAI),\\
1-1 Oho, Tsukuba 305-0801, Japan \\
\email{azumat@post.kek.jp, 
sbal@stp.dias.ie, 
nagao@mx.ibaraki.ac.jp, 
jnishi@post.kek.jp}} 

\abstract{
``Fuzzy CP$^2$'',
which is a four-dimensional fuzzy manifold
analogous to the fuzzy 2-sphere (S$^{2}$),
appears as a classical solution in the dimensionally reduced
8d Yang-Mills model with a cubic term involving the 
structure constant of the SU(3) Lie algebra.
Although the fuzzy S$^2$, which is also a classical 
solution of the same model, has actually smaller free energy
than the fuzzy CP$^2$,
Monte Carlo simulation shows that the fuzzy CP$^2$ is stable
even nonperturbatively due to the suppression of tunneling effects
at large $N$ as far as the coefficient of the cubic term ($\alpha$)
is sufficiently large.
As $\alpha$ is decreased, 
both the fuzzy CP$^2$ and the fuzzy S$^2$
collapse to a solid ball and 
the system is essentially described by the pure Yang-Mills model
($\alpha = 0$). The corresponding transitions are of first order.
The gauge group generated dynamically
above the critical point
turns out to be of rank one for both CP$^2$ and S$^2$ cases.
Above the critical point,
we also perform perturbative calculations 
for various quantities {\em to all orders}, 
taking advantage of the one-loop saturation of the effective action
in the large-$N$ limit.
By extrapolating our Monte Carlo results
to $N=\infty$, we find excellent agreement with the all order results.
}


\newcommand{\bel}{\begin{equation}\label}

\newcommand{\non}{\nonumber \\}
\newcommand{\n}{\nonumber}
\newcommand {\beq}{\begin{equation}}
\newcommand {\eeq}{\end{equation}}
\newcommand {\beqa}{\begin{eqnarray}}
\newcommand {\eeqa}{\end{eqnarray}}
\newcommand {\bc}{\begin{center}}
\newcommand {\ec}{\end{center}}
\newcommand {\tr}{{\rm tr\,}}

\newcommand {\ee}{\mbox{e}}

\newcommand{\A}{{\tilde A}}
\newcommand{\stwo}{{\textrm S}^{2}}

\newcommand{\stwostwo}{{\textrm S}^{2} \times {\textrm S}^{2}}

\def\dag{\dagger}

\def\vs5{\vspace*{5mm}}
\def\vs1{\vspace*{1cm}}
\def\vs2{\vspace*{2cm}}
\def\hs5{\vspace*{5mm}}
\def\hs1{\hspace*{1cm}}
\def\hs2{\hspace*{2cm}}
\def\vs50{\vspace*{50mm}}
\def\vs20{\vspace*{20mm}}

\def\tr{\hbox{tr}}

\begin{document}

\section{Introduction}

Fuzzy spheres \cite{Madore}, 
which are simple compact noncommutative manifolds,
have been recently discussed extensively in the literature.
One of the motivations comes from the general expectation 
that noncommutative geometry provides
a crucial link to string theory and quantum gravity.
Indeed Yang-Mills theories on noncommutative geometry
appear in a certain low energy limit of string theory \cite{Seiberg:1999vs}.
There is also an independent observation that the space-time 
uncertainty relation, which is naturally realized by noncommutative
geometry, can be derived from some general assumptions
on the underlying theory of quantum gravity \cite{gravity}.
Another motivation is to use fuzzy spheres
as a regularization scheme alternative to 
the lattice regularization \cite{Grosse:1995ar}.
Unlike the lattice, fuzzy spheres preserve the continuous symmetries 
of the space-time considered, and hence it is expected that 
the situation concerning chiral symmetry 
\cite{Grosse:1994ed,%
Carow-Watamura:1996wg, chiral_anomaly, non_chi,balagovi,%
chiral_anomaly2,balaGW,Nishimura:2001dq,AIN,AIN2,Ydri:2002nt,%
Iso:2002jc,Balachandran:2003ay,nagaolat03, AIN3,0412052,0509034,0602078,0604093} 
and supersymmetry might be improved.

As expected from the connection to string theory \cite{Myers:1999ps},
fuzzy spheres appear as classical solutions
in matrix models with a Chern-Simons-like term
\cite{0003187,0101102,0204256,0207115,0301055}
and their dynamical properties have been studied
in refs. \cite{0108002,0206075,0303120,0307075,0309082,0312241,0402044,0403242,0506062,0506205}. 
One can actually use matrix models to define a regularized 
field theory on the fuzzy spheres as well as on a noncommutative 
torus \cite{AMNS}, which enables nonperturbative studies of such
theories from first principles \cite{simNC}. 
These matrix models belong to the class of the 
so-called dimensionally reduced models
(or large-$N$ reduced models),
which is widely believed to provide a constructive definition of 
superstring and M theories \cite{9610043,9612115,9703030}.
The space-time is represented by the eigenvalues of the bosonic matrices,
and in the IIB matrix model \cite{9612115}, in particular,
the dynamical generation of {\em four}-dimensional space-time 
(in {\em ten}-dimensional type IIB superstring theory) has been discussed
by many authors \cite{Aoki:1998vn,Ambjorn:2000bf,%
Ambjorn:2000dx,NV,Burda:2000mn,%
Ambjorn:2001xs,exact,sign,Nishimura:2001sx,%
KKKMS,Kawai:2002ub,Vernizzi:2002mu,0307007,Nishimura:2003rj}.

In ref.\ \cite{0401038} we have studied
the dimensionally reduced 3d Yang-Mills-Chern-Simons (YMCS) model, 
which has the fuzzy 2-sphere (S$^{2}$)
as a classical solution \cite{0101102}.
Unlike previous works we have performed nonperturbative first-principle
studies by Monte Carlo simulation.
We observed a first-order phase transition as we vary 
the coefficient ($\alpha$) of the Chern-Simons term.
For small $\alpha$
the large-$N$ behavior of the model is the same as in the 
pure Yang-Mills model, whereas for large $\alpha$
a single fuzzy S$^2$ appears dynamically.

For obvious reasons it is interesting to extend this work to
a matrix model which accommodates a {\em four}-dimensional fuzzy manifold.
In ref. \cite{0405096} 
we have studied the dimensionally reduced 5d Yang-Mills model with 
the quintic Chern-Simons term \cite{0204256,0301055}, 
which is known to have the fuzzy 4-sphere (S$^4$)
as a classical solution \cite{9712105}.
Unlike the fuzzy S$^2$ case, however, the fuzzy S$^{4}$ is unstable at the 
classical level, and Monte Carlo simulation confirmed that it does not
stabilize even at the quantum level.
The negative result is essentially due to the fact that 
the {\em quintic} Chern-Simons term has higher powers in $A_\mu$ 
than the Yang-Mills term.

This motivates us to return to the class of models with a {\em cubic} term.
As a candidate of a four-dimensional fuzzy manifold,
we study the fuzzy CP$^{2}$ \cite{0207115,0309082,%
Grosse:1999ci,Alexanian:2001qj,Karabali:2002im,Carow-Watamura:2004ct},
which appears as a classical solution
in the dimensionally reduced 8d Yang-Mills model with a cubic term
involving the structure constant of the SU(3) Lie algebra.
In fact the fuzzy S$^2$, which is also a classical solution of this model,
has smaller free energy than the fuzzy CP$^{2}$.
Monte Carlo simulation shows, however, that the fuzzy CP$^{2}$ is
stable even nonperturbatively 
due to the suppression of tunneling effects at large $N$
as far as the coefficient ($\alpha$) of the cubic term 
is sufficiently large.
As we decrease $\alpha$, both the fuzzy CP$^{2}$ and the fuzzy S$^2$
collapse to a solid ball, and the system is essentially described by the 
pure Yang-Mills model ($\alpha = 0$).
The corresponding phase transitions are of first order,
and the lower critical point agrees with the analytical result
obtained from the one-loop effective action.
Since the one-loop effective action is saturated at one loop
in the large-$N$ limit, the analytical result for the critical point 
is expected to be free from higher loop corrections.

Above the critical point, we perform perturbation calculations 
for various quantities {\em to all orders}, 
taking advantage of the one-loop saturation of the effective action
in the large-$N$ limit.
This technique was originally proposed for a supersymmetric model
\cite{0403242}, where the effective action is saturated only at two loop.
In the bosonic case, one can obtain all order results by performing
essentially the one-loop calculation \cite{0410263}.
By extrapolating our Monte Carlo results for various observables 
to $N=\infty$, we find excellent agreement with the all order results.

  In the large-$N$ reduced models, not only the space-time \cite{Aoki:1998vn}
  but also the gauge group \cite{9903217}
  is expected to appear dynamically. 
  While there are certain evidences in the IIB matrix model that
  indeed {\em four-dimensional} space-time appears dynamically 
  \cite{Nishimura:2001sx,KKKMS,Kawai:2002ub,0307007,0506033,0510263},
  the issue of the gauge group is totally unclear.
  The models we are studying may be considered as a toy model in which
  one may obtain a definite answer to such a question, since
  the gauge group of rank $k$ naturally
  appears if the true vacuum is given by $k$ coincident fuzzy manifolds.
  The value of $k$ should be determined dynamically, 
  and the result may of course depend on the model one considers.
  In the present model we find the gauge group to be
  of rank one for both the fuzzy CP$^2$ and the fuzzy S$^2$.
  In arriving at this conclusion, the existence of the first-order
  phase transition plays a crucial role as 
  in our previous work \cite{0401038}.

  This paper is organized as follows.
  In section \ref{modeldef} we define the model and show that
  the fuzzy CP$^{2}$ and the fuzzy S$^{2}$ appear as classical solutions.
  In sections \ref{propCP2} and \ref{propS2}
  we study the properties of the fuzzy CP$^{2}$ and the fuzzy S$^{2}$,
  respectively.
  In section \ref{truevac} we compare the free energy
  for the fuzzy CP$^{2}$ and the fuzzy S$^{2}$ to discuss 
  which is the true vacuum whenever they exist.
  In section \ref{k-coincident} we study the properties of the
  $k$ coincident fuzzy CP$^{2}$.
  In section \ref{dyn_gauge} we determine the rank of the dynamical 
  gauge group for the fuzzy CP$^{2}$ and the fuzzy S$^{2}$.
  Section \ref{sum_dis} is devoted to a summary and discussions.
  In appendix \ref{sten} we present
  the explicit form of the fuzzy CP$^2$ configuration.
  In appendix \ref{One-loop-free-energy} 
  we formulate the perturbation theory around fuzzy manifolds
  and obtain an expression for the one-loop free energy.
  In appendices \ref{cp2oneloop} and \ref{s2oneloop}
  we show the perturbative calculations for the fuzzy 
  CP$^{2}$ and the fuzzy S$^{2}$, respectively.

\section{The model and its classical solutions}
\label{modeldef}

  The model we study is defined by the action
  \begin{eqnarray}
   S = N \, \tr \left( - \frac{1}{4} 
\, [A_{\mu},
   A_{\nu}]^{2} + \frac{2}{3}  \, i \, \alpha \, 
  f_{\mu \nu \rho} \, A_{\mu} \, A_{\nu} \, A_{\rho}
  \right) \ ,
  \label{cp2action} 
  \end{eqnarray}
  where $A_{\mu}$ ($\mu = 1, \cdots , 8$) are $N \times N$ traceless
  hermitian matrices.  Here and henceforth we sum over repeated indices.
  The coefficient $f_{\mu \nu \rho}$ is the structure constant 
  of the SU$(3)$ Lie algebra, whose
  nonzero components are given explicitly by
  \begin{eqnarray}
 f_{123} = 1 \, , \quad
   f_{458} = f_{678} = \frac{\sqrt{3}}{2} \, , \quad
f_{147} = f_{246} = f_{257} 
  = f_{345} = f_{516} = f_{637} = \frac{1}{2} \ .
  \label{cp2structure}
  \end{eqnarray}

The pure Yang-Mills model ($\alpha = 0$) and its obvious
generalization to $D$ dimensions with $D$ matrices $A_\mu$ ($\mu=1,
\ldots , D$) have been studied by many authors.  In particular, the
large-$N$ dynamics of the model have been studied by the $1/D$
expansion and Monte Carlo simulation~\cite{9811220}.  The partition
function was conjectured~\cite{Krauth:1998yu} and
proved~\cite{Austing:2001bd} to be finite for $N > D/(D-2)$.  
(See refs.~\cite{Krauth:1998xh,Austing:2001pk,Krauth:1999qw} 
for the supersymmetric case.)  
The partition function in the presence of the Chern-Simons
term has been studied analytically for $N=2$~\cite{0309264}, and it
turned out to be convergent in the supersymmetric case, but not in the
bosonic case.  It is also proved that adding a Myers term (the
cubic term in the present case) does not affect the convergence
as far as the original path integral converges
absolutely~\cite{0310170}, which means, in particular, that the
partition function of our model is convergent for $N\ge 4$.

  The classical equation of motion of the model (\ref{cp2action}) is given by
  \begin{eqnarray}
    [A_{\nu},[A_{\mu}, A_{\nu}]] - i \, \alpha \, f_{\mu \nu \rho} [A_{\nu},
  A_{\rho}] = 0    \ .
  \label{cp2eom}
  \end{eqnarray}
  One can easily see that there exists a solution of the form
  \begin{equation}
     A_{\mu} = \alpha \, T_{\mu} \ , 
  \end{equation}
  where $T_\mu$ satisfies the
  SU$(3)$ Lie algebra
  \begin{eqnarray}
   [T_{\mu}, T_{\nu}] = i f_{\mu \nu \rho} T_{\rho} \ . 
   \label{cp2trepcom}
  \end{eqnarray}
  Hence one obtains a classical solution 
  for each of the $N$-dimensional representations of the SU$(3)$ Lie algebra.
  Let us consider the case in which $T_{\mu}$ is given by
  the irreducible $(m,n)$ representation
  $T^{(m,n)}_{\mu}$. Such a solution exists when the size of the matrices
  $A_\mu$ is
  \begin{eqnarray}
   N = \frac{1}{2} (m+1)(n+1)(m+n+2) \ .
  \end{eqnarray}
  The explicit form of $T^{(m,n)}_{\mu}$ is given in the appendix
  \ref{sten}.

  The space represented by the matrices
  $A_{\mu} = \alpha \, T^{(m,n)}_{\mu}$
  has ${\rm SU}(3)$ isometry. 
  There are two kinds of manifold whose
  isometry is ${\rm SU}(3)$. One is 
  ${\rm SU}(3)/{\rm U}(2)$, and the other is
  ${\rm SU}(3)/({\rm U}(1)\times {\rm U}(1))$. 
  In fact the ${\rm SU}(3)/{\rm U}(2)$ space corresponds 
  to the $(m,0)$ or the $(0,n)$ representation,
  whereas the ${\rm SU}(3)/({\rm U}(1) \times {\rm U}(1))$ 
  space corresponds to the $(m,m)$ representation.

  In what follows we consider the ${\rm CP}^2 = {\rm SU}(3)/{\rm U}(2)$ 
  space, which is given by
  \begin{eqnarray}
   A^{({\rm CP}^2)}_{\mu} \equiv \alpha \, T^{(m,0)}_{\mu} \ .  
   \label{cp2solm0}
  \end{eqnarray} 
  Note that this solution exists only for 
  \begin{eqnarray}
    N = \frac{1}{2}(m+1)(m+2) = 3,6,10,15,21, \cdots \  .
  \end{eqnarray} 
  This is in contrast with the fuzzy S$^{2}$ case \cite{0401038}, 
  where the corresponding solution exists for arbitrary $N$.
  As in ref.\ \cite{0401038} we define the ``radius-squared matrix''
   \begin{eqnarray}
    Q = 
(A_{\mu})^{2} \ ,
    \label{cp2casimir}
   \end{eqnarray}
   which is useful for distinguishing various solutions.
  Using the identity (\ref{casimir}), we obtain
  \begin{eqnarray}
   Q  = \rho ^{2} \, {\bf 1}_{N} \ , \quad \quad
   \rho = \alpha \, \sqrt{\frac{m(m+3)}{3}}  \label{cp2radius}
  \end{eqnarray}
  for the ${\rm CP}^2$ solution, 
  which implies that the fuzzy ${\rm CP}^2$ has the radius $\rho$.

  We note that the model also has the fuzzy S$^2$ type of 
  classical solutions. As an example, let us consider
   \begin{eqnarray}
    A^{({\rm S}^{2})}_{\mu} \equiv
      \left\{ \begin{array}{ll} \alpha \, L_{\mu}^{(N)}  & 
    \textrm{~for~$\mu=1,2,3$} \ ,  \\ 0 & \textrm{~otherwise} \ ,
    \end{array}
      \right. \label{cp2s2sol}
   \end{eqnarray}
   where $L_{\mu}^{(N)}$ is the $N$-dimensional irreducible
   representation of the SU$(2)$ Lie algebra.
   The radius-squared matrix $Q$ is given in this case as
   \begin{eqnarray}
    Q  =  R^{2} \, {\bf 1}_{N} \ , \quad\quad
   R =  \frac{1}{2} \, \alpha \,  \sqrt{N^{2}-1} \ .
   \label{cp2radiuss2}
   \end{eqnarray}
   Although there are other fuzzy S$^2$ solutions with a smaller 
   radius, the one given above is considered 
   the most relevant since it has the smallest action among them.

  \section{Properties of the fuzzy CP$^{2}$}
\label{propCP2}
  In order to study the properties of the fuzzy CP$^{2}$,
  we simulate the model (\ref{cp2action})
  using (\ref{cp2solm0}) as the initial configuration. 
  We apply the heat bath algorithm developed in ref.\ \cite{9811220}
  for the pure Yang-Mills model by
  implementing the cubic term as explained in the appendix A of 
  ref. \cite{0401038}. 
  \FIGURE{
    \epsfig{file=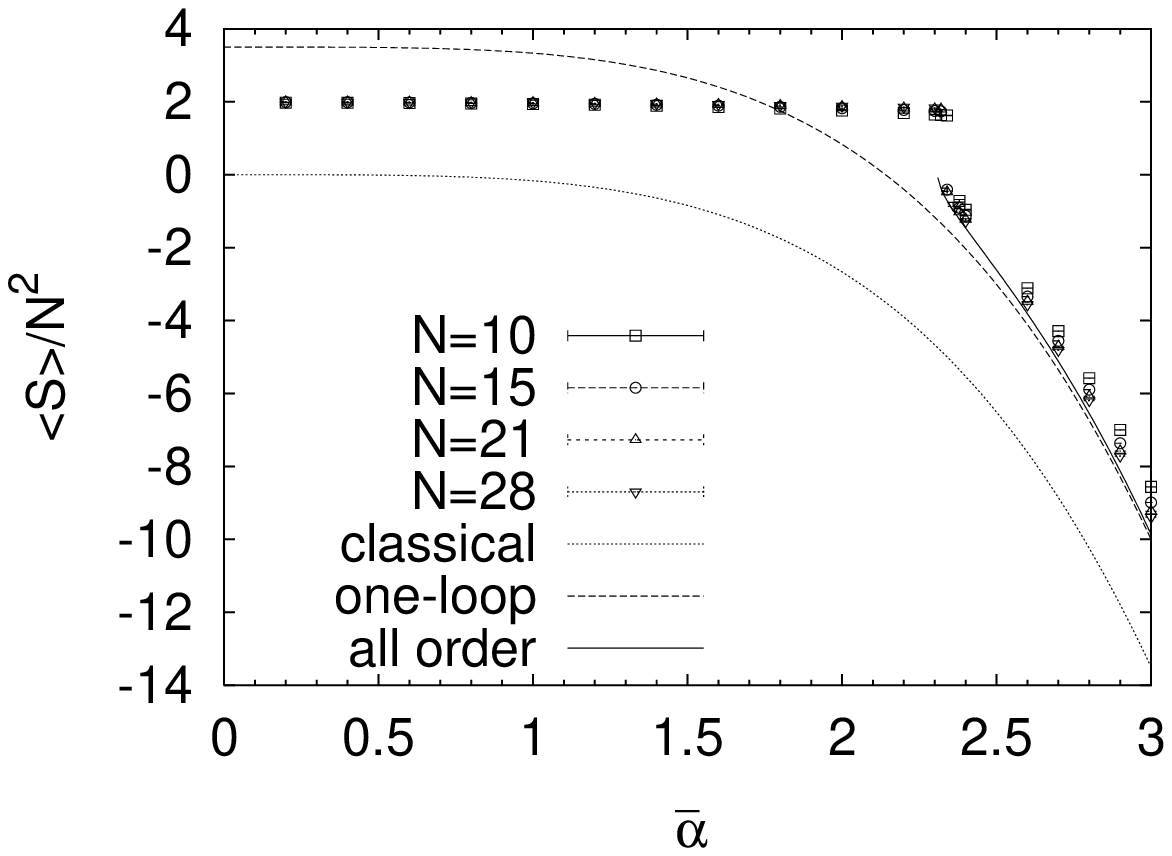,width=7.4cm}
    \epsfig{file=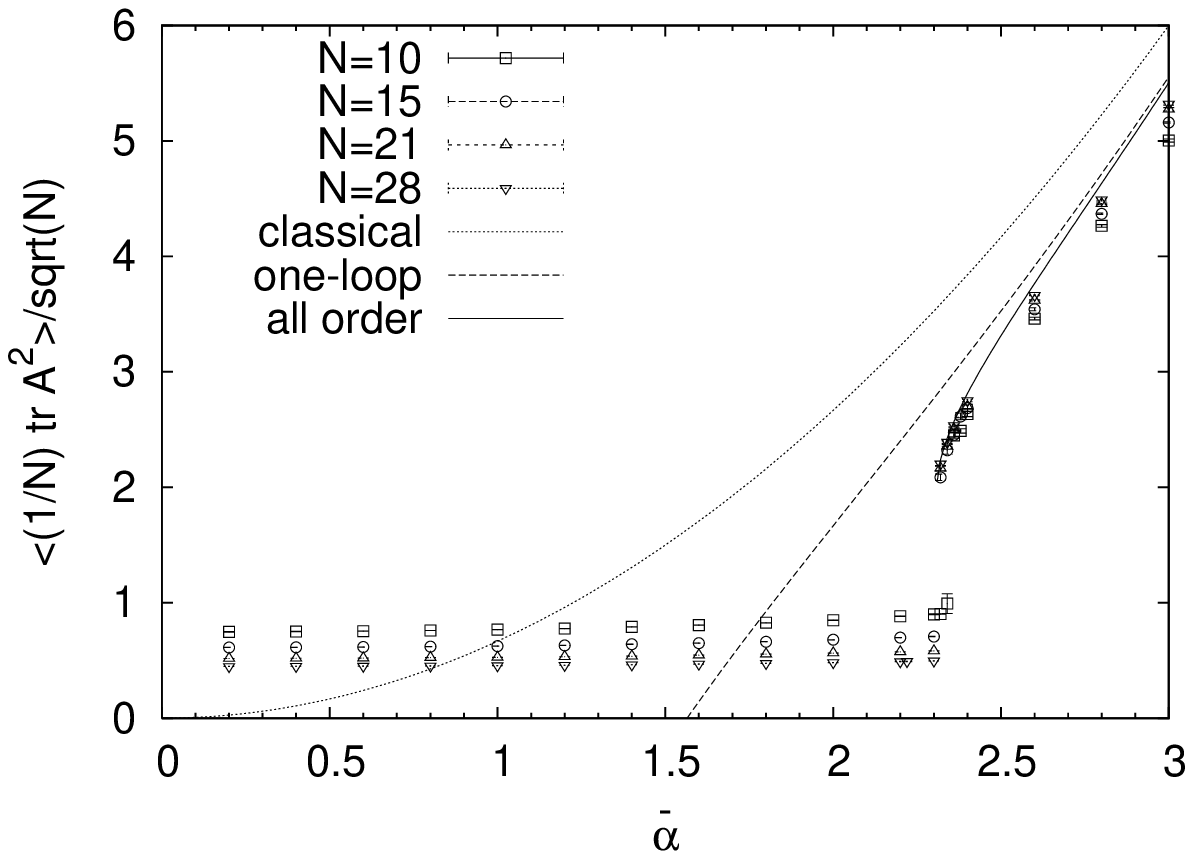,width=7.4cm}
   \caption{The quantities
     $\frac{1}{N^2}
    \langle S \rangle$ and $\frac{1}{\sqrt{N}} \langle \frac{1}{N}
     \tr (A_\mu)^{2} \rangle$ are
     plotted against ${\bar \alpha} = \alpha N^{\frac{1}{4}}$ 
     for $N=10, 15, 21,28$ ($m=3,4,5,6$) for the fuzzy CP$^2$ start.
     The dotted, dashed and solid lines represent 
     the classical, one-loop and all order results, respectively, 
     at large $N$.}
   \label{miscCP2FS}}
 
  \FIGURE{
    \epsfig{file=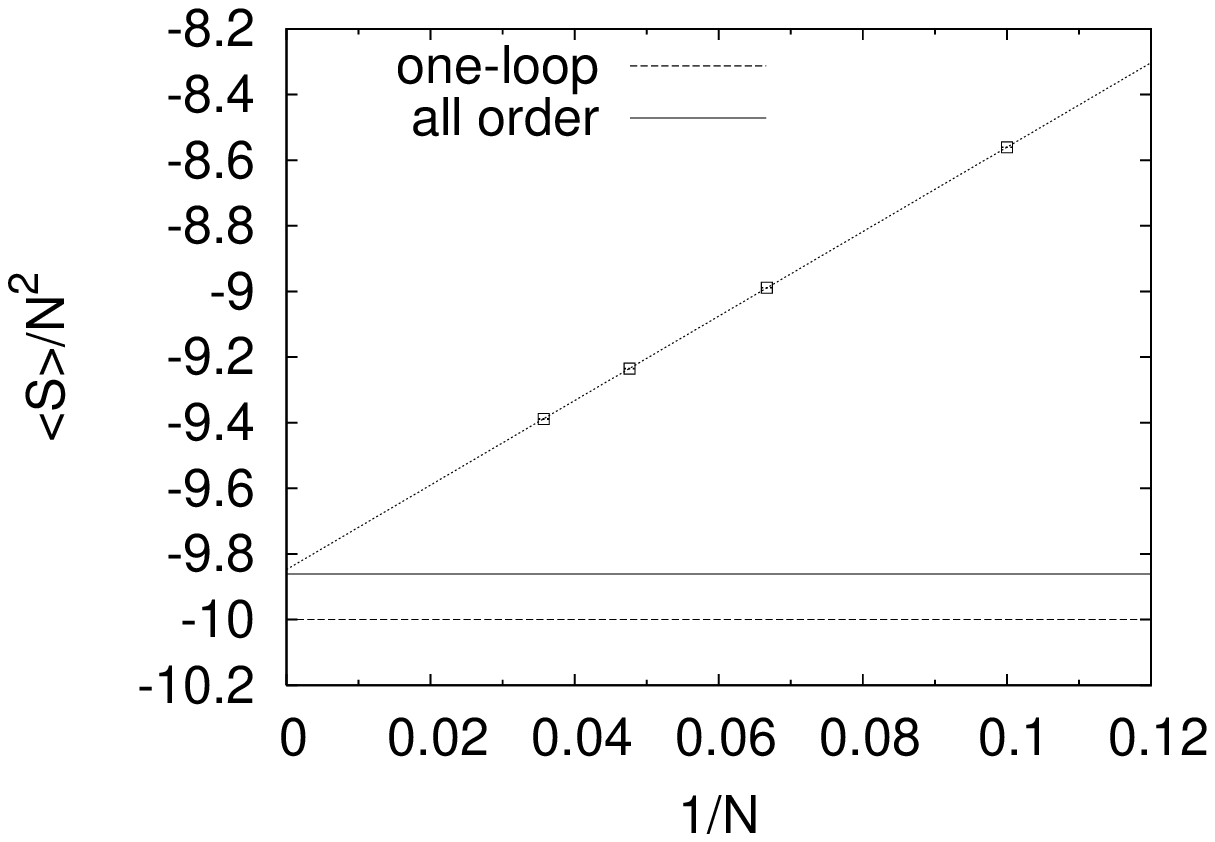,width=7.4cm}
    \epsfig{file=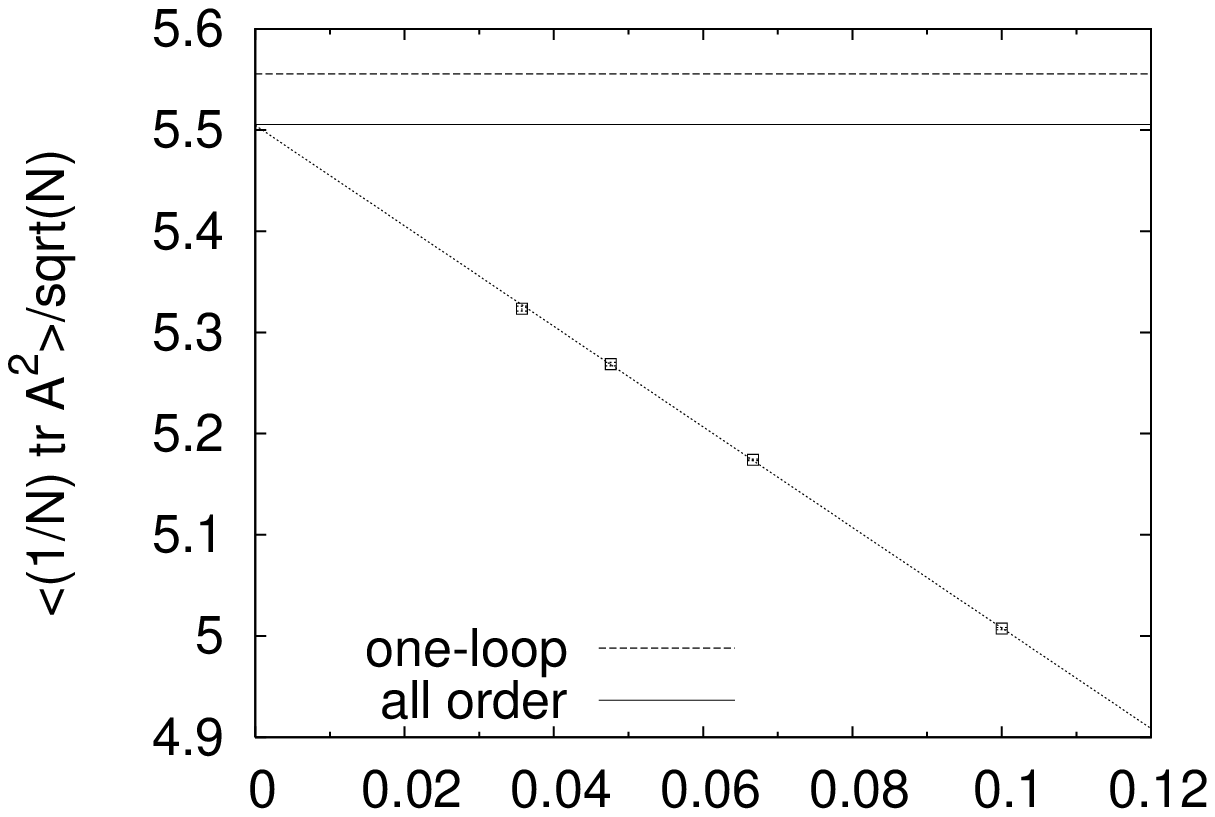,width=7.4cm}
   \caption{The quantities
     $\frac{1}{N^2} \langle S \rangle $ (left) and 
     $\frac{1}{\sqrt{N}}
     \langle \frac{1}{N} \tr (A_{\mu})^{2} \rangle $ (right)
     are plotted against $\frac{1}{N}$ for ${\bar \alpha} = 3.0$
     for the fuzzy CP$^2$ start.
     The dotted lines represent a linear fit, and 
     the horizontal solid (dashed) lines represent 
     the all order (one-loop) results obtained analytically at $N=\infty$.
     The large-$N$ extrapolation demonstrates
     excellent agreement with the all order result.}
   \label{cp2finiteN}}

{}From perturbative calculations, it turns out natural
to fix the rescaled parameter 
 \begin{eqnarray}
  {\bar \alpha} = \alpha \, N^{\frac{1}{4}} \ , \label{cp2rescaled}
 \end{eqnarray}
when we take the large-$N$ limit.
In figure \ref{miscCP2FS}
we plot the Monte Carlo results for the action
$\langle S \rangle$ and the ``extent of space-time''
$\langle \frac{1}{N} \tr (A_\mu)^{2} \rangle$
(with an appropriate normalization factor)
against the rescaled parameter ${\bar \alpha}$ 
for $N=10,15,21,28$ ($m=3,4,5,6$).
 We observe a discontinuity around
  \begin{eqnarray}
   {\bar \alpha} = 
 {\bar \alpha}^{(\rm{CP}^{2})}_{\rm cr} \simeq 2.3 \ , 
 \label{cp2critical}
  \end{eqnarray}
  which suggests the existence of a first-order phase transition. 
  An analogous first-order phase transition has been found also 
  in the 3d YMCS model \cite{0401038}.
  The critical point (\ref{cp2critical}) agrees well with
  the analytical result (\ref{anal_crit_CP2})
with $k=1$.

 Below the critical point, the Monte Carlo results 
 are almost independent of $\alpha$.
 This is because the cubic term in the action (\ref{cp2action})
 takes small values, 
 and hence it does not play any role in this regime. 
 Note, in particular, 
 that $\frac{1}{N} \langle \tr(A_\mu)^2 \rangle  \simeq {\rm O}(1)$
 as in the pure Yang-Mills model ($\alpha = 0$) \cite{9811220},
 which means that
 $\frac{1}{N} \langle \tr(A_\mu)^2 \rangle$ has different
 large-$N$ behaviors in the two phases, as clearly seen in
figure \ref{miscCP2FS}.

  Above the critical point,  
  we observe that Monte Carlo results approach the all order results
 (\ref{acto-exact}), (\ref{a-sq-exact}) with $k=1$
  as $N$ increases. In order to clarify the finite-$N$ effects, 
  in figure \ref{cp2finiteN} we plot the same quantities against
  $\frac{1}{N}$ for $N=10,15,21,28$ ($m=3,4,5,6$)
  with fixed ${\bar \alpha} = 3.0
    > {\bar \alpha}^{({\rm CP}^{2})}_{\rm cr}$. 
  The data can be nicely fitted to a straight line, which implies that the
  leading finite-$N$ effect is of ${\rm O}(\frac{1}{N})$.
  This allows us to make a reliable extrapolation to $N=\infty$,
  and we find excellent agreement with the all order results.

%

 \section{Properties of the fuzzy S$^{2}$}
\label{propS2}

 \FIGURE{
    \epsfig{file=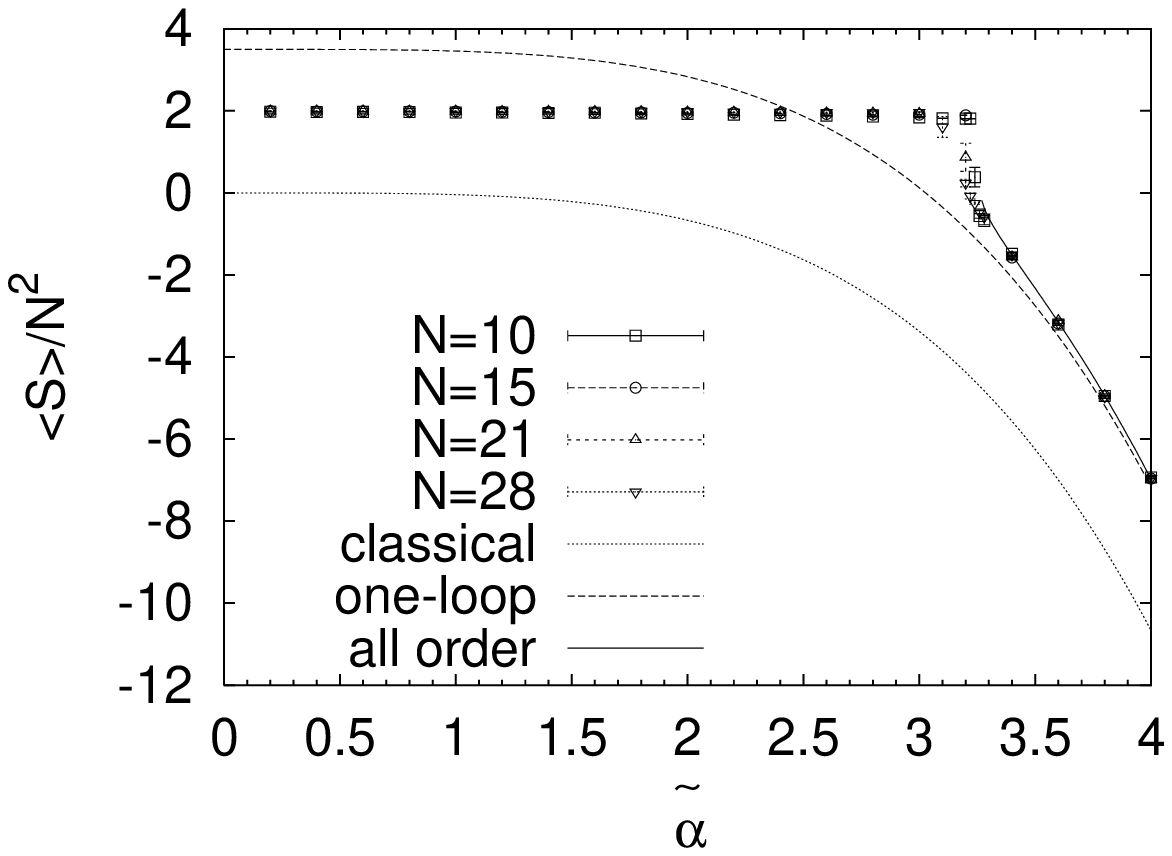,width=7.4cm}
    \epsfig{file=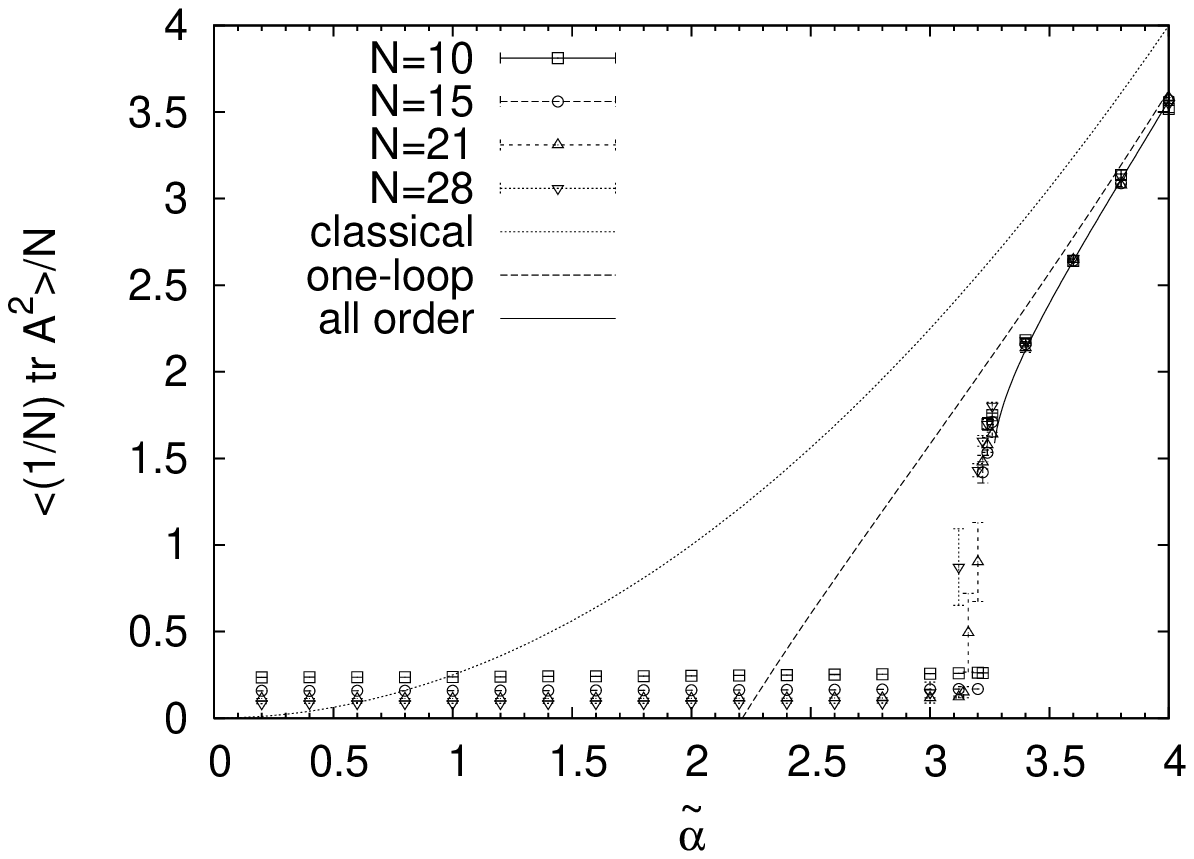,width=7.4cm} 
   \caption{
    The quantities $\frac{1}{N^{2}}
   \langle S \rangle $ and $\frac{1}{N} \langle \frac{1}{N}
    \tr (A_\mu)^{2} \rangle$ are plotted against ${\tilde \alpha} = 
    \alpha N^{\frac{1}{2}}$ for $N=10, 15, 21, 28$
     with the fuzzy $\stwo$ start. The dotted, dashed and solid lines 
     represent the classical, one-loop and all order results, respectively,
     at large $N$.}
   \label{miscCP2S2FS}}

   \FIGURE{
    \epsfig{file=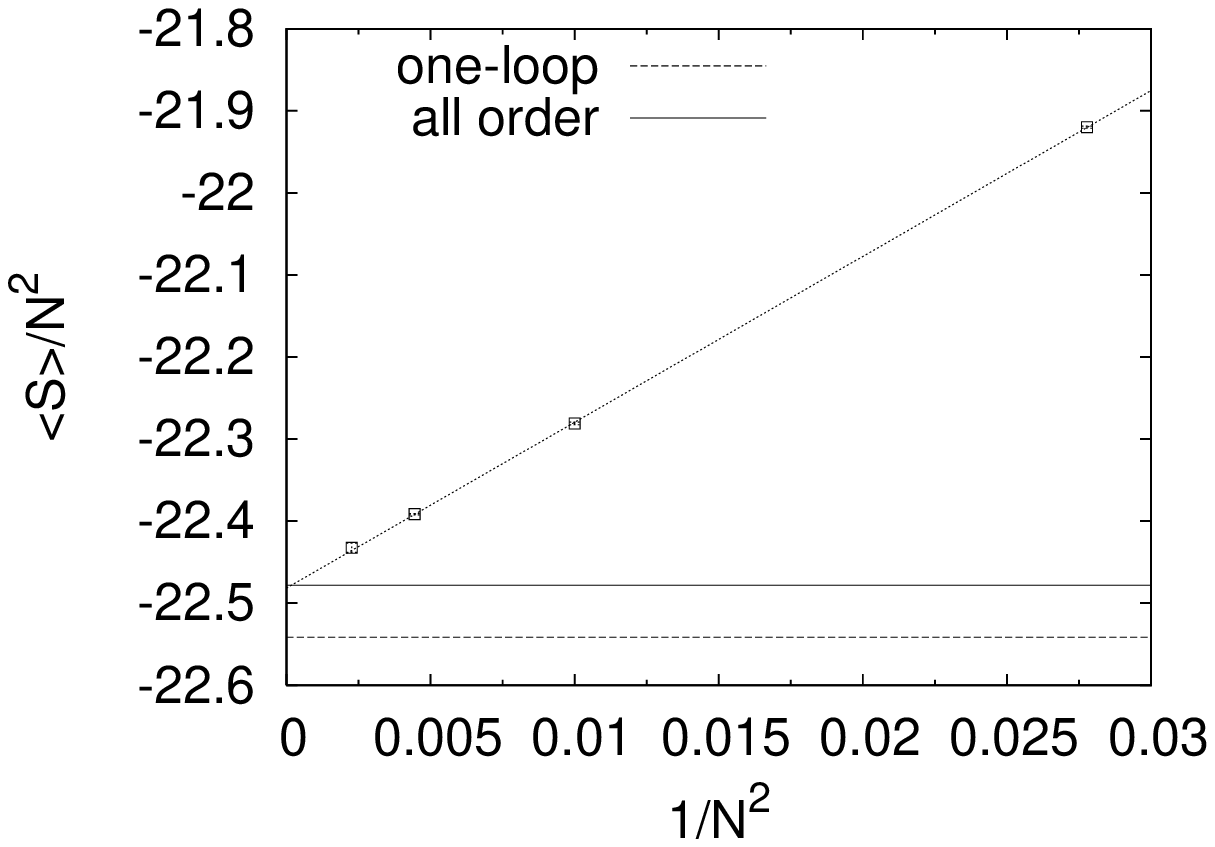,width=7.4cm}
    \epsfig{file=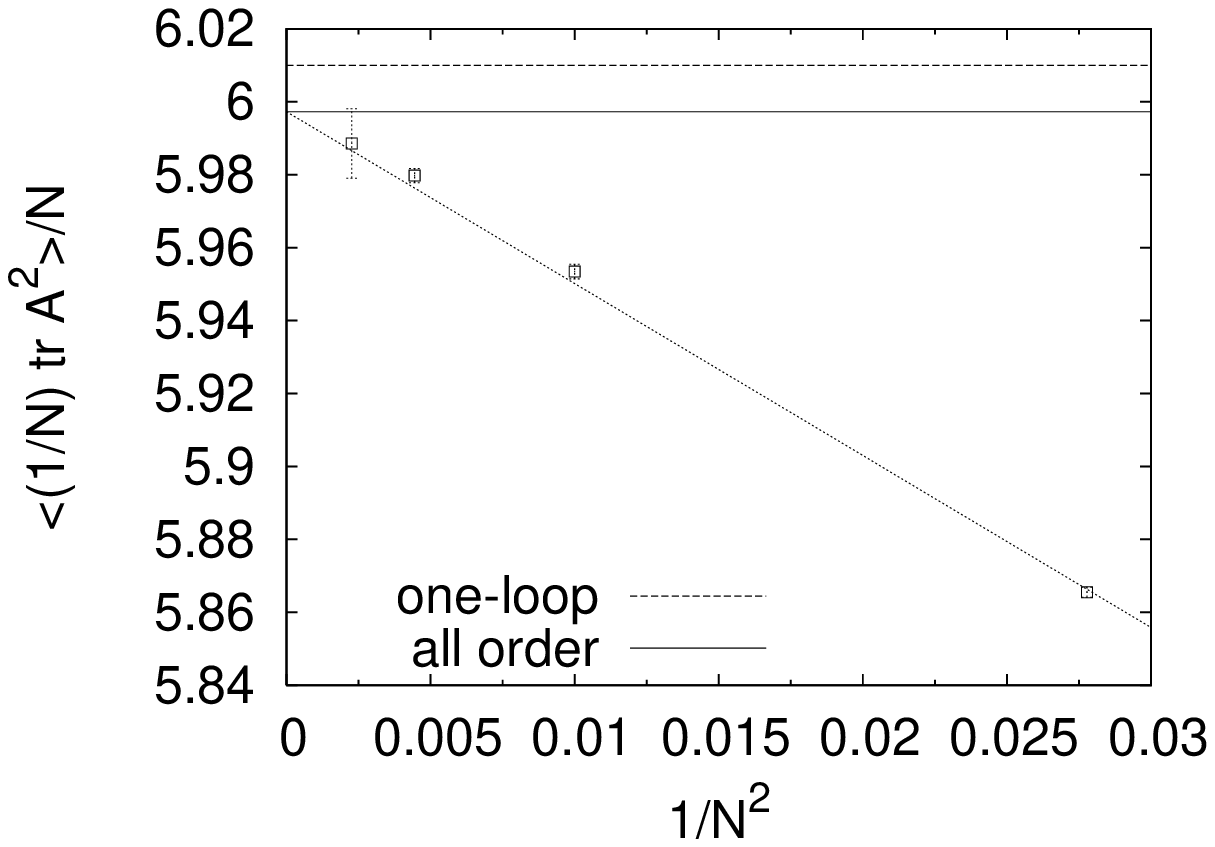,width=7.4cm}
   \caption{The quantities
     $\frac{1}{N^2} \langle S \rangle $ (left) and 
     $\frac{1}{N}\langle
     \frac{1}{N}  \tr (A_{\mu})^{2} \rangle$ (right)
     are plotted against $\frac{1}{N^{2}}$ for ${\tilde \alpha} = 5.0$
     with the fuzzy S$^2$ start.
     The dotted lines represent a linear fit, and 
     the horizontal solid (dashed) lines represent 
     the all order (one-loop) results obtained analytically at $N=\infty$.
     The large-$N$ extrapolation demonstrates
     excellent agreement with the all order result.}
   \label{cp2s2finiteN}}

 In this section we study the properties of the fuzzy S$^2$
 by Monte Carlo simulation using (\ref{cp2s2sol}) as the initial 
 configuration. 
Perturbative calculations suggest that
 the natural definition of the rescaled parameter in this case is
  \begin{eqnarray}
   {\tilde \alpha} = \alpha \, N^{\frac{1}{2}} \ ,
   \label{cp2s2rescaled}
  \end{eqnarray}
 unlike (\ref{cp2rescaled}) in the fuzzy CP$^2$ case.
 In figure \ref{miscCP2S2FS} we plot
 $\frac{1}{N^{2}} \langle S \rangle$ and
 $\frac{1}{N} \langle \frac{1}{N} \tr (A_{\mu})^{2} \rangle$
 against ${\tilde \alpha}$.
 We observe a discontinuity at
 \begin{eqnarray}
  {\tilde \alpha} = {\tilde \alpha}_{\rm cr}^{({\rm S}^{2})} \simeq 3.2  \ ,
  \label{cp2s2critical}
 \end{eqnarray}
 which suggests the existence of a first-order phase transition.
 The critical point (\ref{cp2s2critical}) agrees well with
 the analytical result (\ref{anal-crit_S2}) with $k=1$.
 In terms of the unrescaled parameter $\alpha$,
 the critical point for the 
 fuzzy CP$^2$ and the fuzzy S$^2$ are 
 $\alpha_{\rm cr}^{({\rm CP}^{2})} \simeq \frac{2.3}{N^{1/4}}$ and
 $\alpha_{\rm cr}^{({\rm S}^{2})} \simeq \frac{3.2}{N^{1/2}}$, 
respectively,
 which means that $\alpha_{\rm cr}^{({\rm S}^{2})} <
\alpha_{\rm cr}^{({\rm CP}^{2})}$.
 Below the critical point, 
 Monte Carlo results are identical to those for the fuzzy CP$^2$ start
 presented in the previous section, as expected.

 Above the critical point 
 ${\tilde \alpha} > {\tilde \alpha}^{({\rm S}^{2})}_{\rm cr}$,
 Monte Carlo results are quite close to the all order results 
 (\ref{acto-exact-s2}), (\ref{a-sq-exact-s2}) with $k=1$.
 In figure \ref{cp2s2finiteN} 
 we plot the two quantities against $\frac{1}{N^{2}}$
 for $N=6,10,15,21$ with fixed ${\tilde \alpha} = 5.0
 > {\tilde \alpha}_{\rm cr}^{({\rm S}^{2})}$.
 Monte Carlo results can be nicely fitted to a straight line,
 which implies that the leading finite-$N$ effect is O($\frac{1}{N^2}$),
 as compared with O($\frac{1}{N}$) for the fuzzy CP$^2$ case.
 The large-$N$ extrapolation demonstrates perfect agreement with 
 the all order results obtained in the large-$N$ limit.

 \section{CP$^2$ versus S$^2$ ---which is the true vacuum? ---
} 
\label{truevac}

 In the previous two sections, we have seen that both the fuzzy CP$^2$
 and the fuzzy S$^2$ are stable for sufficiently large $\alpha$.
 In this section we discuss which of the two describes the true 
 vacuum.
 For that purpose we compare the free energy for
the fuzzy ${\rm CP}^2$ and the fuzzy ${\rm S}^2$,
which are obtained to all orders in perturbation theory 
in the large-$N$ limit
as (\ref{w-exact}) and (\ref{w-exact-s2}), respectively.
Setting $k=1$ and rewriting in terms of the unrescaled parameter $\alpha$,
the free energy reads
   \begin{eqnarray}
    W^{({\rm CP}^{2})} 
    &\simeq&  N^{2} \left( - \frac{\alpha^{4} N}{6} + 6 \log 
    \alpha + \log (8 N^7) - \frac{9}{\alpha^{4} N} 
    - \frac{63}{\alpha^{8} N^2}
     - \frac{1485}{2 \alpha^{12} N^3} - \cdots \right), \label{cp2eff1loop} \\
    W^{({\rm S}^{2})} 
    &\simeq & N^{2} \left( - \frac{\alpha^{4} N^{2}}{24} + 6 \log
   \alpha + \log N^{10} - \frac{36}{\alpha^{4} N^{2}} 
   - \frac{1008}{\alpha^{8} N^{4}}
   - \frac{47520}{\alpha^{12} N^{6}} - \cdots \right) \ . \label{s2eff1loop}
  \end{eqnarray} 
 Thus in the region
$
\alpha \ge \alpha_{\rm cr}^{({\rm CP}^{2})} = \frac{2.3}{N^{1/4}} \ ,
$
 where both the fuzzy ${\rm CP}^2$ and the fuzzy ${\rm S}^2$ exist,
 we find that the fuzzy ${\rm S}^2$ has smaller 
 free energy than the fuzzy ${\rm CP}^2$ at large $N$.
Therefore the fuzzy ${\rm CP}^2$ cannot be the 
true vacuum in the present model.

 We note, however, that
 the fuzzy CP$^{2}$ appears to be very stable.
 For instance, we have performed a simulation with
  the fuzzy CP$^{2}$ start for $N=10$ ($m=3$) and $\alpha =
  1.4$, which is just above the critical point.
  The fuzzy CP$^{2}$ does not decay into the fuzzy S$^2$
  even after $10^{7}$ sweeps of the heat bath algorithm. This suggests
  the existence of a potential barrier between the two vacua, which 
  presumably increases with $N$. We therefore consider that
  the fuzzy CP$^{2}$ stabilizes due to the suppression 
  of tunneling effects in the large-$N$ limit.

\section{Properties of the $k$ coincident fuzzy CP$^{2}$}
 \label{k-coincident}

   \FIGURE{
    \epsfig{file=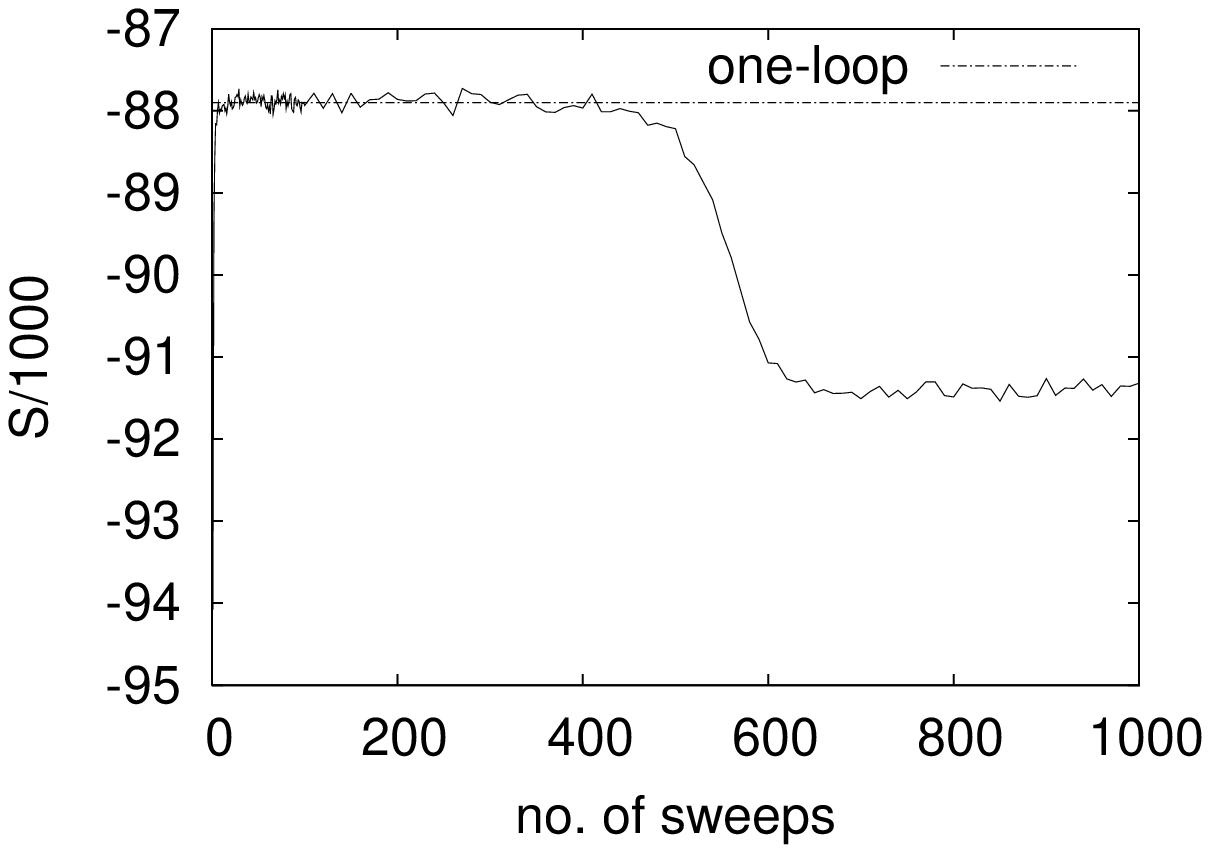,width=7.4cm}
    \epsfig{file=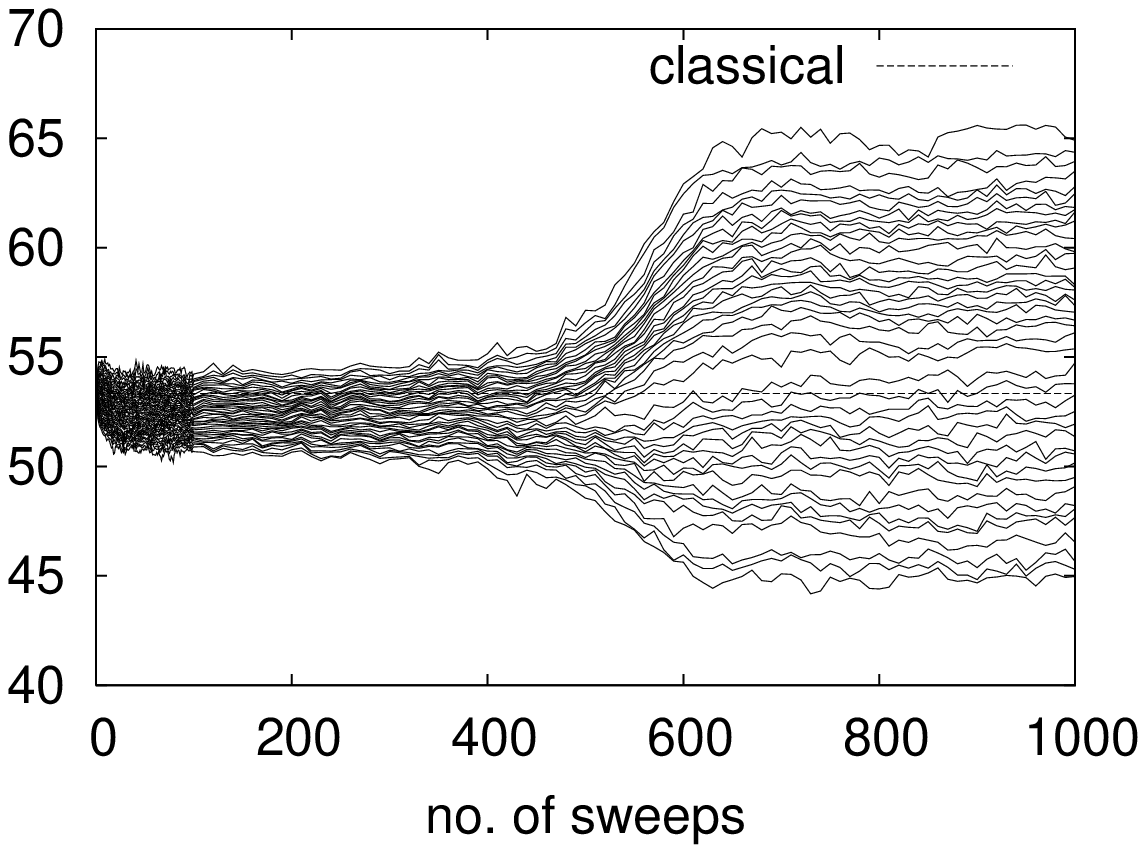,width=7.4cm}
   \caption{The history of the action $S$ (left)
and the eigenvalues of the radius-squared matrix $Q$ (right)
for $\alpha = 2.0$ and $N=42$ using the $k=2$ coincident fuzzy CP$^2$
as the initial configuration. 
The horizontal line in the left (right) plot represents 
the 
one-loop (classical) result for the $k=2$ coincident fuzzy CP$^2$
at $\alpha = 2.0$ and $N=42$.}
   \label{cp2decay}}

   \FIGURE{
    \epsfig{file=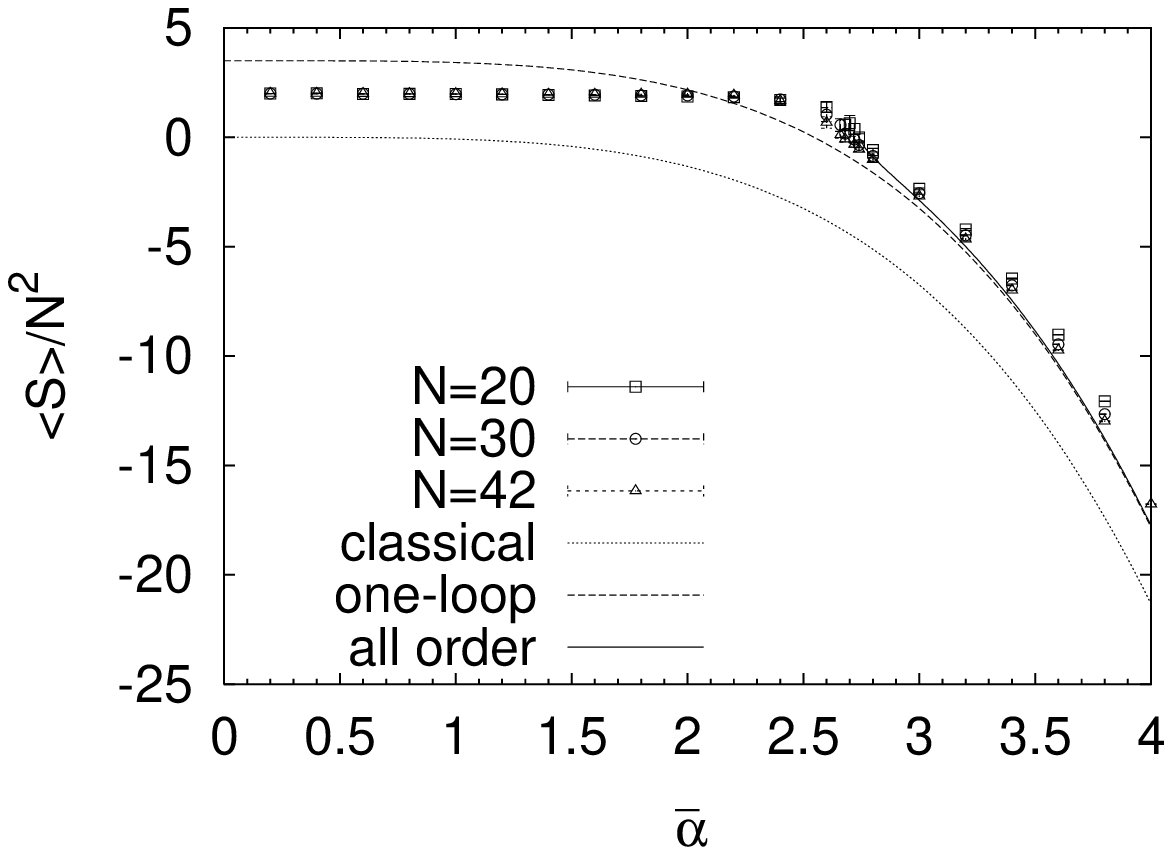,width=7.4cm}
    \epsfig{file=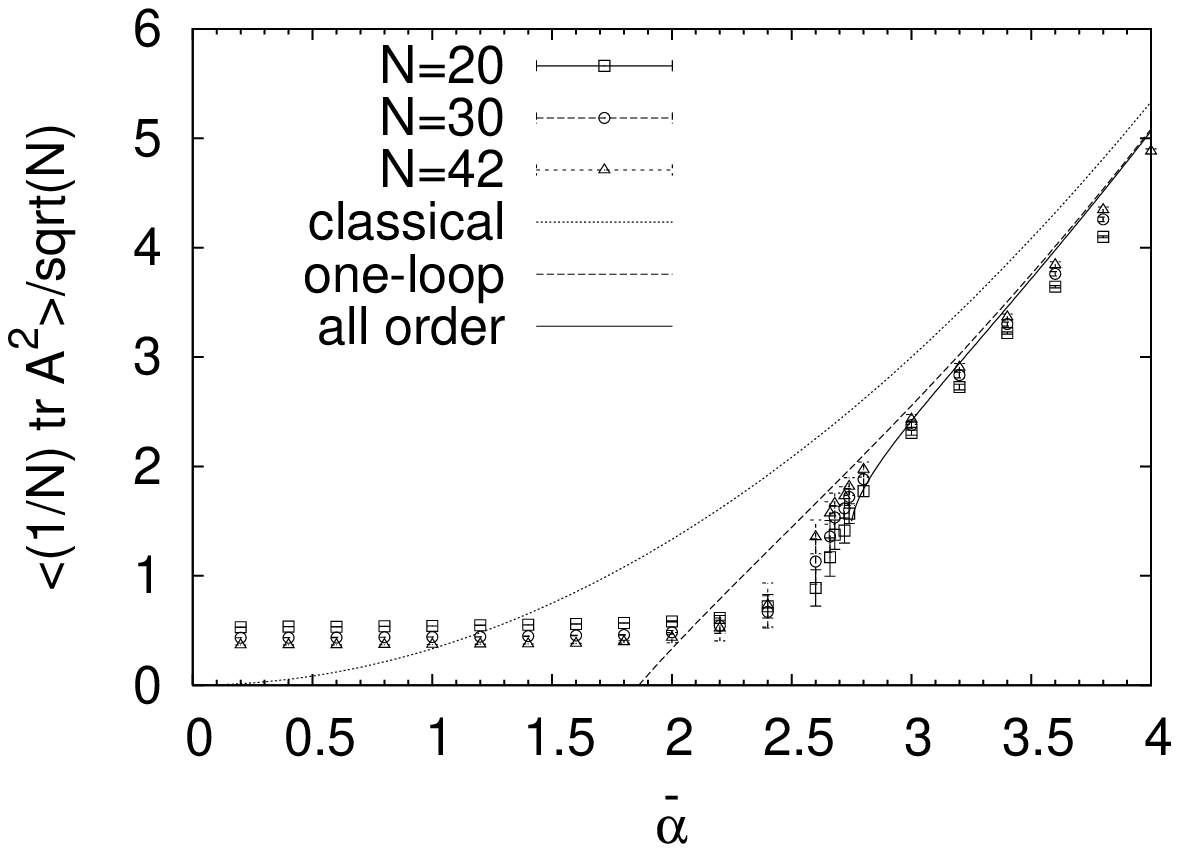,width=7.4cm} 
   \caption{
    The quantities $\frac{1}{N^2}
    \langle S \rangle$ and $\frac{1}{\sqrt{N}} \langle \frac{1}{N}
     \tr (A_\mu)^{2} \rangle$ are
     plotted against ${\bar \alpha} = 
   \alpha N^{\frac{1}{4}}$ for $N=20, 30, 42$
   ($m=3,4,5$) with the $k=2$ coincident fuzzy CP$^2$ start. 
   The dotted, dashed and solid lines represent 
   the classical, one-loop and all order results, respectively,
   at large $N$.}
   \label{miscCP2i002}}

  In this section we discuss the properties of the $k$ coincident
  fuzzy CP$^{2}$ configuration
   \begin{eqnarray}
    A^{(k \, {\rm CP}^2)}_{\mu} 
    \equiv \alpha \,  T^{(m,0)}_{\mu} \otimes {\bf 1}_k \ ,
   \label{cp2k2ini}
   \end{eqnarray}
 which is also a classical solution of the model.
  The size of the matrices should now be
   \begin{eqnarray}
    N = \frac{1}{2} \, k \, (m+1)(m+2) \ . \label{cp2ksize}
   \end{eqnarray}
  Such a configuration is important since it gives rise to a
  gauge theory on the fuzzy CP$^{2}$ with the gauge group of rank $k$.

  We have performed Monte Carlo simulation 
  with the initial configuration given by (\ref{cp2k2ini})
  with $k=2$.
  In figure \ref{cp2decay} we plot the action $S$ and the
  eigenvalues of the radius-squared matrix $Q$ 
  defined by (\ref{cp2casimir})
  against the number of ``sweeps'' 
  in the heat bath algorithm \cite{9811220}.
  We observe that the $k=2$ fuzzy CP$^2$ decays after 600 sweeps.

  Although the $k=2$ coincident fuzzy CP$^2$ is thus only meta-stable,
  we may measure various observables before it actually decays
  in Monte Carlo simulation.
  In figure \ref{miscCP2i002} we plot the results
  against ${\bar \alpha}$ for $N=20, 30, 42$ ($m=3,4,5$).
  We observe a discontinuity at
  \begin{eqnarray}
  {\bar \alpha} = {\bar \alpha}_{\rm cr}
  ^{(k=2 \, {\rm CP}^{2})} \simeq 2.7 \ ,
  \label{cp2criticalk2}
  \end{eqnarray}
  which agrees with the analytical result (\ref{anal_crit_CP2}).
  Above the critical point, our Monte Carlo results agree well with
  the all order results (\ref{acto-exact}), (\ref{a-sq-exact})
  with $k=2$.

 \section{Dynamical gauge group}
 \label{dyn_gauge}

 In the previous section, we have seen in Monte Carlo simulation 
that the $k$ coincident fuzzy CP$^2$ is unstable in the $k=2$ case.
 This instability is related to the zero modes that appear in 
 the perturbation theory around these configurations.
(See appendix \ref{1-loop-free-cp2}.)
 Although we cannot exclude the possibility that the instability disappears
 in the large-$N$ limit, we show that the multi-fuzzy CP$^2$ 
 cannot be the true vacuum anyway, by comparing the free energy
 calculated omitting the zero modes.
 The explicit form of the free energy to all orders in perturbation
 theory is obtained as (\ref{w-exact})
 above the critical point (\ref{anal_crit_CP2}), which we plot
 for $k=1, \cdots, 6$ in figure \ref{free-plot} (left). 
 We find that the $k=1$ case gives the smallest free energy for all values
 of $\bar{\alpha}$.
 Thus, we conclude that
 the dynamical gauge group for the fuzzy CP$^{2}$ is of rank one.
 
 We repeat the same analysis for the fuzzy S$^{2}$ case. 
  The free energy
  for the $k$ coincident fuzzy S$^{2}$ is given 
  to all orders in perturbation theory above the critical point
  (\ref{anal-crit_S2}) by (\ref{w-exact-s2}),
  which we plot for $k=1, \cdots, 6$ in figure \ref{free-plot} (right).
  We find that the $k=1$ case gives the smallest free energy 
  for all values of $\tilde{\alpha}$.
  Therefore, the dynamical gauge group is of rank one in this case as well.

  Let us emphasize that the existence of the first-order phase transition
  plays an important role in arriving at these conclusions.
  As one can see from the free energy (\ref{w-exact}), (\ref{w-exact-s2}),
  the classical term favors small $k$, while 
  the one-loop term proportional to $\log k$ favors large $k$. 
  Therefore, if we disregard 
  the existence of the first-order phase transition, the rank of 
  the dynamical gauge group could be $k > 1$ at small $\alpha$.
  What actually happens is that the critical point (\ref{anal_crit_CP2}),
  (\ref{anal-crit_S2}) increases with $k$, and since we have to 
  restrict ourselves to the region above the critical point, 
  the one-loop term cannot compensate the effect of the classical term.

   \FIGURE{
    \epsfig{file=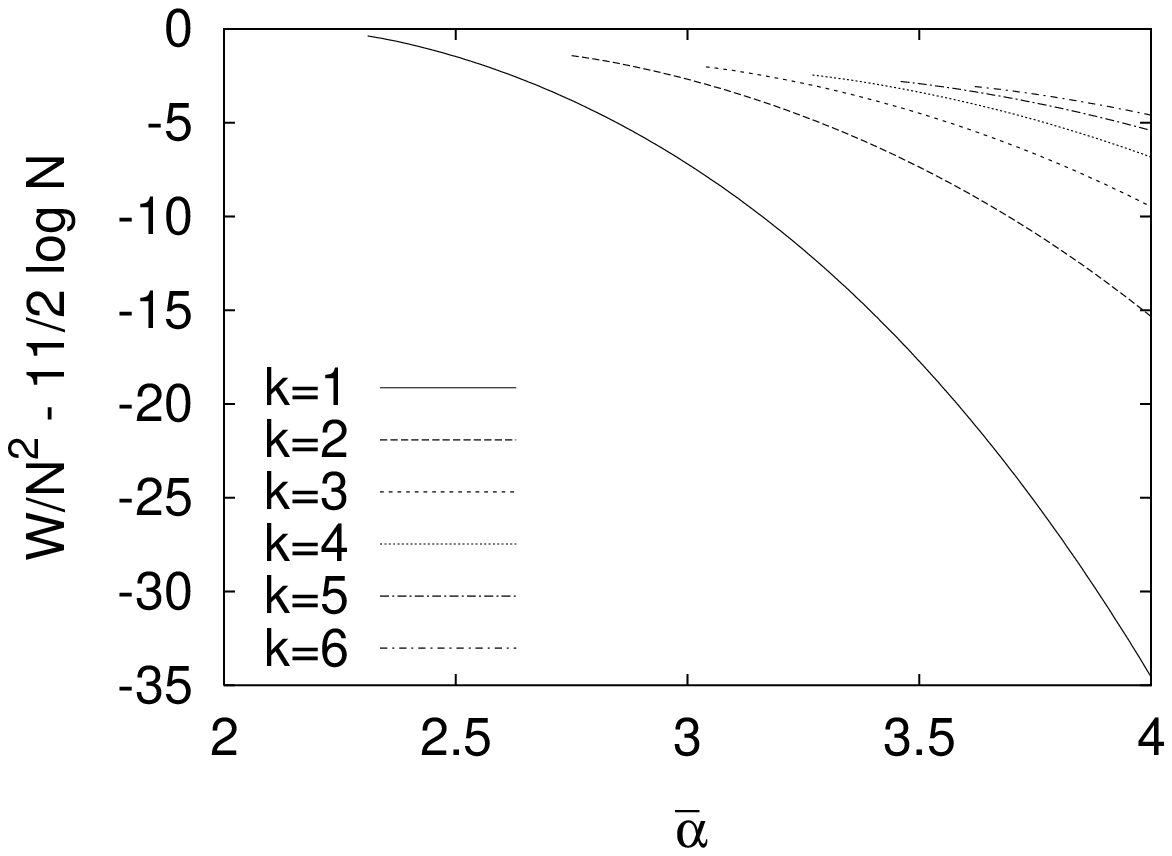,width=7.4cm}
    \epsfig{file=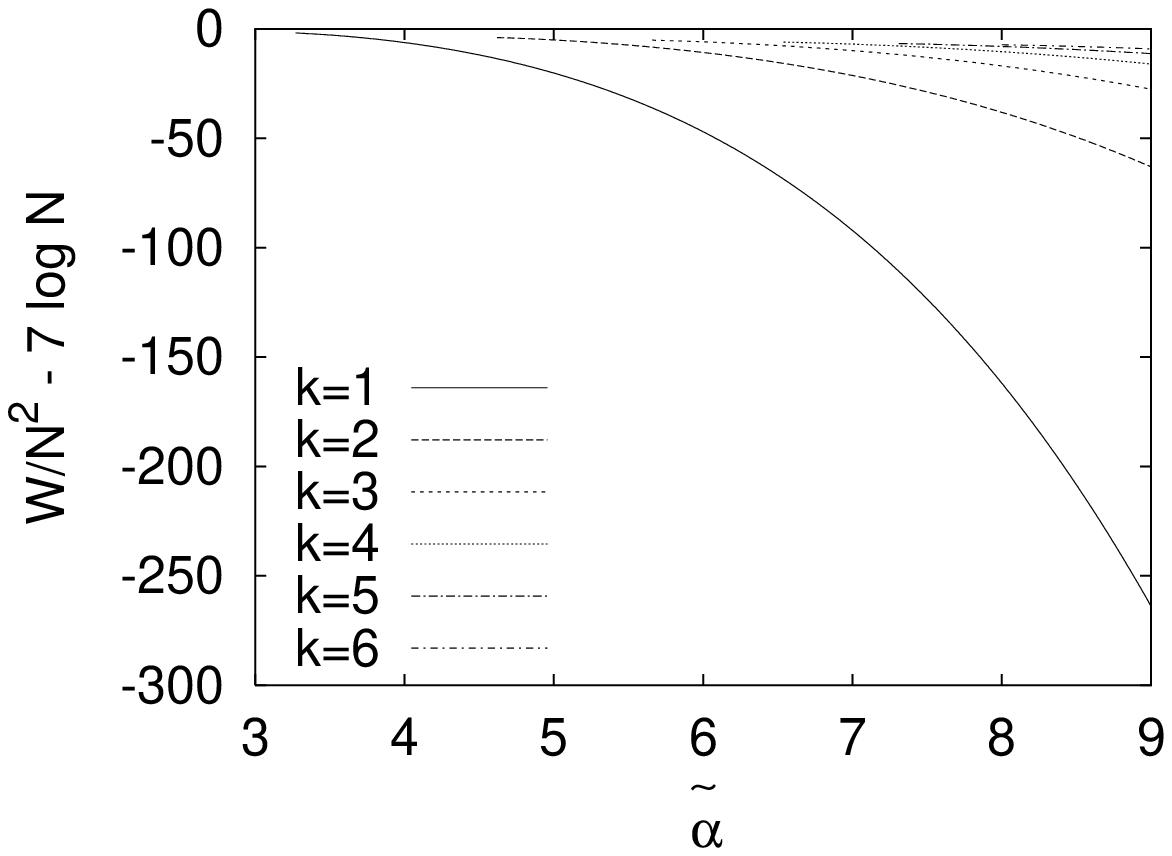,width=7.4cm} 
   \caption{The free energy obtained to all orders in perturbation theory
   is plotted above the critical point for the
   $k$ coincident fuzzy CP$^{2}$ (left) and 
   fuzzy S$^{2}$ (right) with $k=1,\cdots,6$.
   We have taken the large-$N$ limit after subtracting 
   the irrelevant constant term proportional to $\log N$.
   In both cases, the $k=1$ case represented by the solid line
   gives the smallest free energy.}
   \label{free-plot}}

\section{Summary and discussions}
\label{sum_dis}

In this paper we have applied nonperturbative techniques and ideas,
which have been developed in ref.\ \cite{0401038},
to a four-dimensional fuzzy manifold, the fuzzy CP$^2$.
The present model may be considered as a natural extension of our
previous model in the sense that the epsilon tensor, which is 
nothing but the structure constant of SU(2), 
is now replaced by that of SU(3).
Since the number of bosonic matrices should be equal to or larger than the 
dimensionality of the algebra, it is taken to be 8 instead of 3.
Unlike the fuzzy S$^4$ in the matrix model with a quintic term 
\cite{0405096}, the fuzzy CP$^2$ in the present model is nonperturbatively
stable in the large-$N$ limit despite the fact that it has larger
free energy than the fuzzy S$^2$ in the same model.
Thus the model provides a nonperturbative definition of a gauge theory
on the fuzzy CP$^2$.
It would be interesting to investigate the field theoretical aspects
of this model as in the case of noncommutative torus \cite{simNC}.

{}From the viewpoint of the dynamical generation of space-time,
however, we should note that
the fuzzy CP$^2$ cannot be realized as the {\em true vacuum}
since it has larger free energy than the fuzzy S$^2$.
This conclusion is in contrast to the results 
\cite{Nishimura:2001sx,KKKMS,Kawai:2002ub,0307007}
obtained in the IIB matrix model \cite{9612115}, where
four-dimensional space-time is shown to have smaller free energy
than the space-time with other dimensionality.
We should also note that the gauge group dynamically generated
for the cases studied in this paper as well as in the previous
work \cite{0401038} turns out to be of rank one,
although there is no reason for it {\em a priori}.
Indeed ref.\ \cite{0504217} presents an explicit model in which
the gauge group with higher rank is realized in the true vacuum.
In the IIB matrix model, we feel that supersymmetry 
plays an important role in obtaining
four-dimensional space-time as well as a gauge group of sufficiently high
rank that can accommodate the Standard Model.
We would like to report on these issues in the near future.

\acknowledgments
We would like to thank Yusuke Kimura,
Yoshihisa Kitazawa and Dan Tomino for helpful discussions. 
The work of T.A., S.B., K.N.\ and J.N.\
is supported in part by Grant-in-Aid for
Scientific Research (Nos.\ 03740, P02040, 18740127 and 14740163, respectively)
from the Ministry of Education, Culture, Sports, Science and Technology.

\bigskip 

\appendix

\section{Explicit form of the CP$^2$ configuration}
\label{sten}

In this section we present the explicit form of 
the representation matrix $T^{(m,n)}_{\mu}$ of the SU(3) algebra.
This, in particular, provides us with the explicit form
of the fuzzy CP$^2$ configuration $ A^{({\rm CP}^2)}_{\mu}$
in eq.\ (\ref{cp2solm0}).

For that purpose we need to
introduce the so-called (anti-)symmetric tensor product.
Let us denote the matrix element of the matrix $A$ for the orthonormal states $|i \rangle$ 
and $|j \rangle$ as $(A)_{ij} = \langle i| A |j \rangle$. The usual
tensor product is defined by
  \begin{eqnarray}
   \langle i_{1},i_{2} | A \otimes B | j_{1},j_{2} \rangle
 = \langle i_{1} | A | j_{1} \rangle \langle i_{2} | B | j_{2}
 \rangle \ ,
\label{ordinarytensor} 
  \end{eqnarray}
where $  |j_{1}, j_{2} \rangle = |j_{1} \rangle |j_{2} \rangle$.
 The (anti-)symmetric tensor product are defined through
 their matrix elements
  \begin{eqnarray}
&~&  _{\textrm{sym}}\langle i_{1}, i_{2} | A \otimes B | j_{1} j_{2} 
  \rangle_{\textrm{sym}} \ ,  \label{symmetrictensor} \\
&~&  _{\textrm{asym}}\langle i_{1}, i_{2} | A \otimes B | j_{1} j_{2} 
  \rangle_{\textrm{asym}} \ ,  \label{antisymmetrictensor}
  \end{eqnarray}
 where $| j_{1}, j_{2} \rangle_{\textrm{sym}}$
 and $| j_{1}, j_{2} \rangle_{\textrm{asym}}$ are 
 the orthonormal (anti-)symmetrized state defined by
  \begin{eqnarray}
  | j_{1}, j_{2} \rangle_{\textrm{sym}}
 &=& \left\{
\begin{array}{ll}
|j_{1} \rangle |j_{2} \rangle & \mbox{~~~~~for~}j_1 = j_2  \ , \\
\frac{1}{\sqrt{2}} ( |j_{1} \rangle |j_{2} \rangle + |j_{2} \rangle |j_{1}
  \rangle)   & \mbox{~~~~~for~}j_1 \neq j_2 \ ,  
\end{array} \right.
   \label{symmetricstate} \\
   |j_{1}, j_{2} \rangle_{\textrm{asym}} &=& \frac{1}{\sqrt{2}}
    (|j_{1} \rangle |j_{2} \rangle - |j_{2} \rangle |j_{1} \rangle) 
\quad\quad \mbox{for~}j_1 \neq j_2 \ ,
   \label{antisymmetricstate}
  \end{eqnarray}
  respectively. The size of the matrices representing the
  symmetric tensor product $(A \otimes B)_{\textrm{sym}}$
  and the anti-symmetric tensor product
  $(A \otimes B)_{\textrm{asym}}$ is $ \ {}_{k+1} {\rm C}_2 \ $ and
  $ \ {}_{k} {\rm C}_2 \ $, respectively, where $k$ is the size of
  the matrices $A$ and $B$.
  The (anti-)symmetric tensor product can be generalized straightforwardly
  to a product of more than two matrices.
  Then the representation matrix of the $(m,0)$ representation is given as
\beq
   T^{(m,0)}_{\mu} 
 = \sum_{j=1}^{m}
\Bigl( 
\underbrace{{\bf 1}_{3 } \otimes \cdots \otimes {\bf 1}_{3 }}_{j-1} \otimes
t_{\mu}  \otimes \underbrace{ {\bf 1}_{3 } \otimes \cdots
       \otimes {\bf 1}_{3 }}_{m-j} 
\Bigr)_{\textrm{sym}} \ ,
\label{m0rep}
\eeq
where $t_{\mu}$ denotes the fundamental $(1,0)$ representation 
of the SU$(3)$ Lie algebra, which is given explicitly as
  \begin{eqnarray}
   & & t_{1} = \frac{1}{2} \left( \begin{array}{ccc} 0 & 1 & 0 \\ 1 &
  0 &
   0 \\ 0 & 0 & 0 \end{array} \right), \hspace{2mm}
   t_{2} = \frac{1}{2} \left( \begin{array}{ccc} 0 & -i & 0 \\ i & 0 &
   0 \\ 0 & 0 & 0 \end{array} \right), \hspace{2mm}
   t_{3} = \frac{1}{2} \left( \begin{array}{ccc} 1 & 0 & 0 \\ 0 & -1 & 
   0 \\ 0 & 0 & 0 \end{array} \right), \hspace{2mm}
   t_{4} = \frac{1}{2} \left( \begin{array}{ccc} 0 & 0 & 1 \\ 0 & 0 &
   0 \\ 1 & 0 & 0 \end{array} \right), \nonumber \\
 & & t_{5} = \frac{1}{2} \left( \begin{array}{ccc} 0 & 0 & -i \\ 0 & 0 &
   0 \\ i & 0 & 0 \end{array} \right), \hspace{2mm}
   t_{6} = \frac{1}{2} \left( \begin{array}{ccc} 0 & 0 & 0 \\ 0 & 0 &
   1 \\ 0 & 1 & 0 \end{array} \right), \hspace{2mm}
   t_{7} = \frac{1}{2} \left( \begin{array}{ccc} 0 & 0 & 0 \\ 0 & 0 & 
   -i \\ 0 & i & 0 \end{array} \right), \hspace{2mm}
   t_{8} = \frac{1}{2 \sqrt{3}} \left( \begin{array}{ccc} 1 & 0 & 0 \\
  0 & 1 &
   0 \\ 0 & 0 & -2 \end{array} \right) \ . \nonumber
  \end{eqnarray}
 The $(0,n)$ representation can be obtained by simply replacing the
 fundamental representation $t_\mu$ by the anti-fundamental representation,
 $s_{\mu} = - t^{\ast}_{\mu}$, in the above definition.

 In order to obtain the $(m,n)$ representation,
 we have to define an orthonormal state 
  $|j_{1},  \cdots , j_{m} ; k_{1},  \cdots , k_{n}
   \rangle_{\textrm{mix}}$ 
 with a mixed symmetry such that it is symmetric with respect to 
 the first $m$ indices and the last $n$ indices, separately,
 and anti-symmetric with respect to the exchange of one of 
 the first $m$ indices and one of the last $n$ indices.
 (This symmetry is exactly the symmetry of the Young tableaux 
 for the $(m,n)$ representation.)
 We denote the tensor product defined with these states as
 $(A_1 \otimes \cdots \otimes A_m \otimes B_1 \otimes \cdots
\otimes B_n )_{\textrm{mix}}$.
 Then the representation matrix of the $(m,n)$ representation is given as
  \begin{eqnarray}
   T^{(m,n)}_{\mu} &=&
\Bigl( 
 \sum_{j=1}^{m} 
\underbrace{{\bf 1}_{3 } \otimes \cdots \otimes {\bf 1}_{3 }}_{j-1} \otimes
t_{\mu}  \otimes \underbrace{ {\bf 1}_{3 } \otimes \cdots
       \otimes {\bf 1}_{3 }}_{m-j} \otimes 
 \underbrace{ {\bf 1}_{3 } \otimes \cdots
       \otimes {\bf 1}_{3 }}_{n}  \nonumber \\
&~& + \sum_{k=1}^{n}
 \underbrace{ {\bf 1}_{3 } \otimes \cdots
       \otimes {\bf 1}_{3 }}_{m}\otimes 
\underbrace{{\bf 1}_{3 } \otimes \cdots \otimes {\bf 1}_{3 }}_{k-1} \otimes
s_{\mu}  \otimes \underbrace{ {\bf 1}_{3 } \otimes \cdots
       \otimes {\bf 1}_{3 }}_{n-k} 
\Bigr)_{\textrm{mix}} \ .
\label{sympro}
 \end{eqnarray}
Note also that
  \begin{eqnarray}
\left( T^{(m,n)}_{\mu} \right)^{2} 
 & =&  
\left( 
  \underbrace{((t_{\mu})^{2} \otimes {\bf 1}_3 \otimes 
\cdots \otimes {\bf 1}_3)
  + \cdots }_{m\textrm{ terms}} 
 + \underbrace{({\bf 1}_3 \otimes \cdots \otimes {\bf 1}_3 
\otimes (s_{\mu})^{2}  ) + \cdots }_{n \textrm{ terms}} \right.  \nonumber \\
 &~&  + \underbrace{(t_{\mu} \otimes t_{\mu} \otimes {\bf 1}_3 \otimes \cdots
 \otimes {\bf 1}_3) + \cdots}_{m(m-1) \textrm{ terms}} 
 +  \underbrace{({\bf 1}_3 \otimes \cdots \otimes {\bf 1}_3 
\otimes s_{\mu} \otimes s_{\mu}) +
 \cdots }_{n(n-1) \textrm{ terms}} \nonumber \\
  &~& \left.  + \underbrace{(t_{\mu} \otimes
 \cdots \otimes {\bf 1}_3 \otimes s_{\mu} \otimes \cdots \otimes {\bf 1}_3) + \cdots}_{2mn
 \textrm{ terms}} \right)_{\textrm{mix}}  \nonumber \\
  &=& \left( \frac{4}{3} (m+n) + \frac{1}{3} (m(m-1) + n(n-1)) +
 \frac{1}{6} \times 2mn \right) {\bf 1}_{N} \nonumber \\
  &=&  \frac{m(m+3) + n(n+3) + mn}{3} \, {\bf 1}_{N} \ ,
  \label{casimir} 
  \end{eqnarray}
where we have used the formulae
  \begin{eqnarray}
(t_{\mu} \otimes t_{\mu})_{\textrm{sym}}
   &=& \frac{1}{3} \,  ({\bf 1}_3 \otimes {\bf 1}_3)_{\textrm{sym}}  \ , \\
(t_{\mu} \otimes s_{\mu})_{\textrm{asym}}
   &=& \frac{1}{6} \, ({\bf 1}_3 \otimes {\bf 1}_3)_{\textrm{asym}} \ ,  \\
(t_{\mu})^{2} &=& 
(s_{\mu})^{2} = \frac{4}{3} \, {\bf 1}_{3} \ .
  \label{cp2tensor}
  \end{eqnarray}


\section{Perturbative expansion around fuzzy manifolds}
\label{One-loop-free-energy}

In this section we formulate the perturbation theory
around the classical solution $X_\mu$ 
given either by (\ref{cp2k2ini}) representing
$k$ coincident fuzzy ${\rm CP}^2$
or by (\ref{k-multi-fuzzyS2}) representing
$k$ coincident fuzzy S$^2$.
The single fuzzy ${\rm CP}^2$ and the single fuzzy S$^2$
are included as a special case $k=1$.

Let us evaluate the partition function 
$Z = \int d A \, e^{-S}$ around the classical solution 
$A_{\mu} = X_{\mu}$ at the one-loop level.
We define the measure of the
path integral as
\begin{eqnarray}
 d A = \prod_{\mu=1}^{8} \prod_{a=1}^{N^{2}-1} d A_{\mu}^{a} \ ,
\end{eqnarray}
where $A_{\mu} = \sum_{a=1}^{N^{2}-1} A^{a}_{\mu} \, t^{a}$
with $t^{a}$ being the generators of
SU$(N)$ normalized as tr$(t^{a} t^{b}) = \delta_{ab}$.
We need to fix the gauge since there are flat directions corresponding to
the transformation 
\begin{eqnarray}
 A_{\mu} \to A_{\mu}^{g} = g \, A_{\mu} \, g^{\dagger} \ ,
\end{eqnarray}
where $g$ is an element of the coset space $H = {\rm U}(N) / {\rm U}(k)$.
In order to remove the associated zero modes, 
we introduce the gauge fixing term
and the corresponding ghost term
\begin{eqnarray}
S_{\rm g.f.} &=& -\frac{N}{2}\,
\text{tr} \, [X_\mu, A_\mu]^2 \ , \\
S_{\rm ghost}&=& -N \, 
\text{tr} \Bigl([X_\mu,\bar{c}][A_\mu,c] \Bigr) \ ,
\end{eqnarray}
where $c$ and $\bar c$ are the ghost and anti-ghost fields, respectively.
These ghost fields take values in the tangent space of $H$.
We perform the integration over $A_{\mu}$ by decomposing it into
the classical background and the fluctuation as 
$A_{\mu} = X_{\mu} + \A_{\mu}$.
The partition function can be rewritten as
\begin{eqnarray}
 Z = {\rm vol}(H) \, {\cal N} \int d {\tilde A} \, dc \, d {\bar c} 
 \, e^{-S_{\rm total}} \ ,
 \label{Z-calculated}
\end{eqnarray}
where the total action is defined by
\begin{eqnarray}
S_{\rm total} &=& S + S_{\rm g.f.} + S_{\rm ghost} \ ,
\end{eqnarray}
and it is given explicitly as
\begin{eqnarray}
S_{\rm total} &=& S[X]+ S_{\rm kin} + S_{\rm int}  \ , \\
S_{\rm kin}&=&\frac{1}{2}\,  N \, \tr \left( 
\tilde A_\mu [X_\lambda , [X_\lambda , \tilde A_\mu ]] \right)
+ N \, \tr \Bigl( \bar c \, [X_\lambda , [X_\lambda , c ]] \Bigr) \ , 
\label{SUalg_term_CP2} \\
S_{\rm int}&=&-N \, \tr \left( 
[\tilde A_{\mu} , \tilde A_{\nu} ][X_{\mu} ,\tilde A_{\nu} ] \right) 
- \frac{1}{4}\,  N\, 
\tr \left( [\tilde A_{\mu} , \tilde A_{\nu} ]^2 \right) \non
&&+ \frac{2}{3}\, i \, \alpha f_{\mu \nu \rho} \, N \,
 \tr \left( \tilde A_{\mu}  \tilde A_{\nu}  \tilde A_{\rho}  \right)
+ N \, \tr \left( \bar c \, [X_{\mu}, [\tilde A_{\mu} , c]] \right) \ .
\end{eqnarray}
The normalization factor ${\cal N} = (2 \pi N)^{- (N^{2} - k^{2})/2}$ 
in (\ref{Z-calculated}) can be
obtained by following the usual 
gauge fixing procedure as in ref. \cite{0504217}.
The linear terms in ${\tilde A}_{\mu}$ cancel since $X_\mu$ is assumed to
satisfy the classical equation of motion.  
Since the classical solution $X_\mu$ we are considering
is proportional to $\alpha$,
we can rescale the matrices as $A_\mu \mapsto \alpha \, A_\mu$,
$c \mapsto \alpha \, c$, $\bar c \mapsto \alpha \, \bar c$, 
so that all the terms in the total action $S_{\rm total}$ become
proportional to $\alpha ^{4}$.
This means that the expansion parameter of the present perturbation
theory is $\frac{1}{\alpha ^{4}}$.
The volume of the coset space $H$ in (\ref{Z-calculated}), 
${\rm vol}(H) = {\rm vol}({\rm U}(N))/ {\rm vol}({\rm U}(k))$, 
can be obtained by using the formula
\begin{eqnarray}
 {\rm vol}({\rm U}(p)) = 
\frac{(2 \pi)^{p(p+1)/2}}{(p-1)! \, (p-2)! \cdots 1! \, 0!} \ . 
\end{eqnarray}

We calculate the free energy $W = - \log Z$ as a perturbative expansion
$W = \sum_{j=0}^{\infty} W_{j}$, where $W_{j} = {\rm O}(\alpha^{4(1-j)})$
comes from the $j$-loop contribution.
The classical part is obtained as 
$ W_0 = S[X]$, which is nothing but the action evaluated 
at the classical solution $A_{\mu} = X_{\mu}$.

Introducing
the operator ${\cal P}_{\mu}$
\begin{eqnarray}
 {\cal P}_{\mu} M \stackrel{\rm def}{=} [X_{\mu}, M] \ ,
\end{eqnarray}
which acts on a $N \times N$ traceless hermitian matrix $M$,
the kinetic term (\ref{SUalg_term_CP2}) reads
\begin{eqnarray}
 S_{\rm kin} = N \, \tr \left\{
 \frac{1}{2} \A_{\mu} ({\cal P}_{\lambda})^{2} \A_{\mu}
 + {\bar c} ({\cal P}_{\lambda})^{2} c \right\} \ .
\end{eqnarray}
The one-loop term can therefore be obtained as
\begin{eqnarray}
 W_{1} = 3 \, {\cal T}r \log \{N ({\cal P}_{\lambda})^{2} \} 
 - \log \{ {\rm vol}(H) {\cal N} \} \ ,
\label{W1-def}
\end{eqnarray}
where the symbol ${\cal T}r$ denotes 
the trace in the space of traceless hermitian matrices.

\section{Perturbative calculations for the fuzzy ${\rm CP}^2$}
\label{cp2oneloop}

In this section we focus on the case where
the classical solution $X_\mu$ is taken to
be the $k$ coincident fuzzy ${\rm CP}^{2}$ 
(\ref{cp2k2ini}).
The results for the single fuzzy ${\rm CP}^2$
can be readily obtained by setting $k=1$.

\subsection{One-loop calculation of free energy}
\label{1-loop-free-cp2}

Let us calculate the free energy up to one-loop.
The classical part is obtained as
\begin{equation}
W_0 = -\frac{1}{6} N^2 \alpha^4 (n-1) \ ,
\label{multi_cl_action_CP2}
\end{equation}
where we have defined
\beq
n \equiv  \frac{N}{k} = \frac{1}{2} \, (m+1)(m+2) \ ,
\label{cp2k-n}
\eeq
and used the relation
\begin{equation}
f_{\mu\nu\rho}\, f_{\mu\nu\rho'}=3 \, \delta_{\rho\rho'} \ .
\end{equation}

Next we evaluate the one-loop contribution $W_{1}$
in (\ref{W1-def}). 
In order to solve the eigenvalue problem of the operator 
$({\cal P}_\lambda)^2$,
we introduce an analog of matrix spherical harmonics
in the fuzzy S$^2$ case \cite{0401038}.
Let us denote it as $\{ Y_{st} \}$, 
where the indices $s$ and $t$ run over $s=0,1,\cdots,m$ 
and $t=1,\cdots,(s+1)^{3}$, respectively.
It gives a complete basis for the space of $n\times n$ matrices.
(Note, for instance, $\sum _{s=0}^{m} (s+1)^3 = n^2$.)
For a given $s$, $Y_{st}$ 
transforms as a $(s,s)$-type irreducible 
representation of SU$(3)$ under 
the adjoint operation $[ T_\mu ^{(m,0)} , \ \  \cdot \ \ ]$.
For more details of $Y_{st}$, see
refs.\ \cite{Grosse:1999ci,Alexanian:2001qj,Karabali:2002im,0207115,%
Carow-Watamura:2004ct}.
We also introduce $k \times k$ matrices ${\bf e}^{(a,b)}$, whose 
$(a,b)$ element is 1 and all the other elements are zero.
Then, as a complete basis of $N \times N$ matrices,
we define
\begin{equation}
{\cal Y}_{st}^{(a,b)} \equiv 
Y_{st} \otimes {\bf e}^{(a,b)} \ ,
\end{equation}
which satisfies the relation
\begin{eqnarray}
 \tr \Bigl( {\cal Y}_{st}^{(a,b)\dagger} {\cal Y}_{s't'}^{(a',b')} \Bigr) 
 = \delta_{ss'} \delta_{tt'}  \delta_{aa'} \delta_{bb'} \ .
\end{eqnarray}
The eigenvalue problem of the operator $({\cal P}_{\lambda})^{2}$
can be solved as
\begin{equation}
({\cal P}_{\lambda})^2 {\cal Y}_{st}^{(a,b)}=
\alpha^2 s \, (s+2) {\cal Y}_{st}^{(a,b)} \ .
\end{equation}
Note that ${\cal Y}_{00}^{(a,b)}$ 
for all the $(a,b)$ blocks are the zero modes.
In the $k=1$ case, the zero mode should be excluded since $A_{\mu}$
are traceless. 
For $k \ge 2$ the tracelessness condition 
removes only one of the $k^{2}$ zero modes.
Here we omit the rest of them by hand
\footnote{
In fact these zero modes are responsible for the instability 
of the multi-fuzzy ${\rm CP}^2$
discussed in section \ref{k-coincident}. (See appendix D of ref.\
\cite{0401038} for more in-depth 
discussions on this point in an analogous model.)
However, the number of zero modes, $k^2 -1$, is negligible
compared with the dimension of the configuration space, 
which is of O($N^2$). 
Indeed figure \ref{miscCP2i002} shows that the results
obtained by omitting the zero modes are in reasonable agreement
with Monte Carlo results obtained 
before the decay of the multi-fuzzy ${\rm CP}^2$ 
actually takes place.
}.
Then the one-loop contribution $W_1$ is given by 
\begin{eqnarray}
W_1 = 3 k^2 \sum_{s=1}^{m} (s+1)^3 
\log \left[ N \alpha^2 s(s+2) \right] - 
\log \{ {\rm vol}(H) {\cal N} \} \ . \label{coin_ol_CP2}
\end{eqnarray}

The one-loop free energy is obtained 
at large $N$ as 
\begin{equation}
W_0  + W_1 \simeq N^2 \left(
-\frac{{\bar\alpha}^4}{6k}  +6\log{\bar\alpha}
+ \log \frac{8N^\frac{11}{2}}{k^{3}} 
- \frac{9}{4} \right) \ ,
\label{OL-effact_CP2}
\end{equation}
where the rescaled parameter $\bar{\alpha}$
is defined by (\ref{cp2rescaled}).

\subsection{Derivation of the critical point}
\label{der_crit}

In section \ref{propCP2} we observed a phase transition
in Monte Carlo simulation starting from the fuzzy CP$^2$
configuration. We can derive the critical point 
 in the same way as in the 3d YMCS model \cite{0401038}.
 Let us consider the effective action for 
 a one-parameter family of configurations 
$A_{\mu} = \beta \, T^{(m,0)}_{\mu} \otimes {\bf 1}_{k}$.
 At the one-loop level, it is obtained at large $N$ as
  \begin{eqnarray}
 \Gamma_{\rm 1-loop} (\bar \beta)
  \simeq N^{2} \left\{ \frac{2}{3k}
   \left(\frac{3 {\bar \beta}^{4}}{4} 
  - {\bar \alpha} {\bar \beta}^{3} \right)  +
   6 \log {\bar \beta} 
+ \log \frac{8N^{\frac{11}{2}}}{k^3}
   - \frac{9}{4} \right\}  \ ,
  \label{eff-cp2-1loop}
  \end{eqnarray}
where ${\bar \beta} = \beta N^{\frac{1}{4}}$.  
 The function of $\bar{\beta}$ on the right hand side
has a local minimum for
\beq
{\bar \alpha} >
 {\bar \alpha}_{\rm cr}^{(k {\rm CP}^{2})} = \frac{4}{\sqrt{3}} \, 
  k^{\frac{1}{4}} = 2.3094011 \cdots \times  k^{\frac{1}{4}} \ ,
  \label{anal_crit_CP2}
\eeq
 which determines the (lower) critical point of the 
 first-order phase transition.
 In fact it is known that the effective action is saturated 
 at one loop in the large-$N$ limit in the case of 
 fuzzy $\stwo$ or fuzzy $\stwostwo$ \cite{0303120,0307007}.
 This is the case also for the fuzzy ${\rm CP}^{2}$.
 Therefore, the critical point obtained above is expected to
 be correct to all orders in perturbation theory.



\subsection{One-loop calculation of various observables}
\label{cp2_obs}

In this section we extend the perturbative calculation
to various observables
including those which are studied 
by Monte Carlo simulation in section \ref{propCP2} and
\ref{k-coincident}.
The zero modes, which appear for $k \ge 2$,
is omitted as in the evaluation of the free energy
given in appendix \ref{1-loop-free-cp2}.
We note that the number of loops in the relevant diagrams
can be less than the order of 
$\frac{1}{\alpha^4}$ in the perturbative expansion
since we are expanding the theory around a nontrivial background.
At the one-loop level, the only nontrivial task is to 
evaluate the tadpole $\langle (\tilde A_\mu)_{ij} \rangle$ 
explicitly.

\subsubsection{Propagators and the tadpole}

The propagators for $\tilde A_\mu$ and the ghosts
are given as
\begin{eqnarray}
\left\langle (\tilde A_\mu)_{ij} (\tilde A_\nu)_{kl}
\right\rangle _0
&=& 
\delta_{\mu\nu} \sum_{ab}
\sum_{s=1}^{m}
\sum_{t=1}^{(s+1)^{3}} 
\frac{1}
{N \alpha^2 s(s +2) }
\left( {\cal Y}_{st}^{(a,b)}  \right)_{ij}
\left( {{\cal Y}_{st}^{(a,b)}}^\dag \right)_{kl} , \\
\Bigl \langle(c)_{ij} (\bar c)_{kl} \Bigr\rangle_0 &=& 
\sum_{ab}
\sum_{s=1}^{m} 
\sum_{t=1}^{(s+1)^{3}}
\frac{1}
{N \alpha^2 s(s +2)}
\left( {\cal Y}_{st}^{(a,b)}  \right)_{ij}
\left( {{\cal Y}_{st}^{(a,b)}}^\dag  \right)_{kl} \ , 
\end{eqnarray}
where the symbol $\langle \ \cdot \ \rangle_0$ refers to the
expectation value calculated using the kinetic term $S_{\rm kin}$ 
in (\ref{SUalg_term_CP2}) only.


Due to the symmetries, the tadpole
$\langle \tilde A_\mu \rangle$ 
can be expressed as 
\begin{equation}
\langle\tilde A_\mu \rangle = c \,  X_\mu 
\end{equation}
with some coefficient $c$. Using the identity
\begin{eqnarray}
\tr \left( X_\mu \langle \tilde A_\mu \rangle \right) &=& 
c \, \tr (X_\mu X_\mu) \n \\
&=& \frac{cN}{3} \alpha^2 m(m+3 ) \ ,
\label{inner-product_CP2} 
\end{eqnarray}
the coefficient $c$ can be determined by
calculating the left hand side of (\ref{inner-product_CP2}).

At the leading order in $\frac{1}{\alpha^4}$,
we have
\begin{eqnarray}
\frac{1}{N} \tr
\left( X_\mu \langle \tilde A_\mu \rangle_{\rm 1-loop} \right)
&=& 
 \left\langle \tr(X_\mu \tilde A_\mu ) \, 
\tr \left([\tilde A_{\nu} ,\tilde A_{\rho} ][X_{\nu} ,\tilde
A_{\rho} ] \right) \right\rangle_0 \n \\
&~& 
- \left\langle \tr(X_\mu \tilde A_\mu)
 \,  \tr\left( \frac{2}{3}\, i \, \alpha 
f_{\nu \rho \sigma} \tilde A_{\nu}  \tilde A_{\rho}  \tilde
A_{\sigma}  \right)\right\rangle_0 \n \\
&~&  -  \left\langle \tr(X_\mu \tilde A_\mu)  \, \tr
\left( \bar c \, [X_{\nu} ,[\tilde A_{\nu} ,c] ] \right) 
\right\rangle_0 \ .
\label{tadpole_CP2}
\end{eqnarray}
Using the fact that $X_\mu$ is a linear combination of 
$(Y_{s=1,t} \otimes {\bf 1}_{k})$, 
we can calculate (\ref{tadpole_CP2}) similarly
to the previous section.
After some algebra we arrive at
\begin{equation}
\tr\left(X_\mu\langle\tilde A_\mu \rangle_{\rm 1-loop}\right)
=
-  \frac{2k^2}{ N \alpha^2}(n^2-1) \ .
\label{tr_L^red_A_CP2}
\end{equation}
Using (\ref{inner-product_CP2}) we obtain
\begin{eqnarray} 
\langle\tilde A_\mu \rangle_{\rm 1-loop} 
=
-\frac{6}{\alpha^4} \frac{n^2-1}{n^2 m(m+3)}
\,  X_\mu  
\simeq -\frac{3}{n \alpha^4} \,  X_\mu  \ .
\end{eqnarray}


\subsubsection{One-loop results for various observables}
Using the propagator and the tadpole obtained in the previous 
section, we can evaluate various observables easily at the one-loop
level.

The ``extent of space-time''
$\langle \frac{1}{N} \tr(A_\mu)^2 \rangle$
can be evaluated as
\begin{eqnarray}
\left\langle \frac{1}{N} \tr (A_\mu)^2 \right\rangle
_{\rm 1-loop}
&=&\frac{1}{N}\left[\tr(X_\mu X_\mu)
+  2 \, \tr \left( X_\mu \langle\tilde A_\mu\rangle _{\rm 1-loop} \right)
+\langle\tr (\tilde A_\mu)^2 \rangle_0 \right] \n \\
&=&
\alpha^2 \, \left[
\frac{1}{3} m(m+3) 
-\frac{4}{\alpha^4} \frac{n^2-1}{n^2}
+\frac{8}{n^2 \alpha^4}\sum_{s=1}^{m}
\frac{(s+1)^3}{ s(s+2)} \right] \ . 
\label{tr_A^2_CP2} 
\end{eqnarray}
At large $N$ with fixed $\bar \alpha$, we obtain
\begin{equation}
\frac{1}{\sqrt{N}}
\left\langle \frac{1}{N} \tr(A_\mu)^2 \right\rangle_{\rm 1-loop}
\simeq \frac{2 {\bar \alpha}^{2}}{3k} -\frac{4}{\bar{\alpha}^2} \ . 
\label{tr_A^2_largeN_CP2}
\end{equation}
The expectation value of the Chern-Simons term 
\begin{equation}
M = \frac{2 \, i}{3N} 
   f_{\mu \nu \rho} \tr(A_{\mu} A_{\nu} A_{\rho})
\end{equation}
can be evaluated as
\begin{eqnarray}
\langle M \rangle _{\rm 1-loop}
&=& \frac{2i}{3N}  f_{\mu\nu\rho}
\left[  
\tr(X_\mu X_\nu X_\rho)
+3\, \tr \left( X_\mu X_\nu \langle 
\tilde A_\rho \rangle_{\rm 1-loop} \right) 
\right] \n \\
&=&
-\frac{\alpha^3}{3}m(m+3) + \frac{6(n^2-1)}{\alpha n^2} \ .
\label{o2_CP2}
\end{eqnarray}
At large $N$ with fixed $\bar \alpha$, 
we get
\begin{equation}
\frac{1}{N^\frac{1}{4}} \langle M \rangle _{\rm 1-loop}
\simeq
-\frac{2 \bar{\alpha}^3}{3 \, k^2} 
+ \frac{6}{\bar{\alpha}} \ . 
\label{o2_largeN_CP2}
\end{equation}
The observable 
$\langle \frac{1}{N} \tr (F_{\mu\nu}) ^2 \rangle$
can be calculated in a similar manner, but it is 
easier to make use of the Schwinger-Dyson equation
\begin{equation} 
\left \langle  \frac{1}{N} \tr (F_{\mu\nu})^2 
+ 3\alpha M \right \rangle =
8 \left( 1-\frac{1}{N^2} \right) \ , \label{sde-exact}
\end{equation}
from which we obtain
\begin{eqnarray}
\left \langle \frac{1}{N} \tr 
(F_{\mu\nu})^2 \right \rangle _{\rm 1-loop}
= 
8 \left( 1-\frac{1}{N^{2}} \right)
- 3 \alpha \langle M \rangle _{\rm 1-loop} 
\simeq \frac{2 \bar{\alpha}^4}{k}- 10 \ . 
\label{o1_CP2}
\end{eqnarray}
Combining (\ref{o2_CP2}) and (\ref{o1_CP2}), 
we get 
\begin{eqnarray}
\frac{1}{N^2} \langle S \rangle _{\rm 1-loop} =
\frac{1}{4} 
\left \langle \frac{1}{N} \tr (F_{\mu\nu})^2
\right\rangle _{\rm 1-loop}
+  \alpha \, \langle M \rangle _{\rm 1-loop}
\simeq -\frac{\bar{\alpha}^4}{6 \, k} + \frac{7}{2} \ . 
\label{appendix_(S)_CP2}
\end{eqnarray}

\subsubsection{An alternative derivation}

Since $\tr (F_{\mu\nu})^2$ and $M$ are the quantities 
that appear in the action $S$, we can obtain their expectation values
easily by using the free energy (\ref{OL-effact_CP2})
calculated for the $k$ coincident fuzzy ${\rm CP}^2$.
Let us consider the action
\begin{eqnarray}
   S(\beta_{1}, \beta_{2}; \alpha)
 = N \tr \left( - \frac{\beta_{1}}{4} [A_{\mu},
   A_{\nu}]^{2} \right)
+ \beta_2 N^2 \alpha M \ ,     \label{verydefinition2_CP2}
\end{eqnarray}
where we have introduced two free parameters
$\beta_{1}$ and $\beta_{2}$,
and define the corresponding free energy by
\begin{eqnarray}
  \ee ^ {- W(\beta_{1}, \beta_{2}; \alpha)} 
 =  \int d A \, 
\ee ^{-S(\beta_{1}, \beta_{2}; \alpha)} \ .
\label{oneloopefb12_CP2}
\end{eqnarray}
By rescaling the integration variables as
$A_{\mu} \mapsto \beta_{1}^{- \frac{1}{4}}  A_{\mu}$,
we find
\begin{equation}
   W(\beta_{1}, \beta_{2} ; \alpha )
= 2(N^{2}-1) \log \beta_{1} + 
W(1,1 ;\alpha \beta_{2} \beta_{1}^{-\frac{3}{4}} )   \ .
\end{equation}
Then $\langle \tr (F_{\mu\nu})^2 \rangle$,
$\langle M \rangle$ and $\langle S \rangle$ can be obtained by
\begin{eqnarray}
\left\langle \frac{1}{N} \tr (F_{\mu \nu})^2 \right\rangle &=& 
\frac{4}{N^{2}} \left. 
\frac{\partial W(\beta_{1}, \beta_{2}; \alpha)}{\partial \beta_{1}} 
\right|_{\beta_{1} = \beta_{2} = 1} = 8 
\left(1 - \frac{1}{N^{2}} \right) - \frac{3 {\bar \alpha}}{N^{2}}
 \frac{\partial W(1,1;\alpha)}{\partial {\bar \alpha}} \ , \nonumber \\
\label{one-loopf-sq2_CP2}  \\
\langle M \rangle &=& \frac{1}{\alpha N^{2}} 
\left. \frac{\partial W(\beta_{1},\beta_{2};\alpha)}{\partial \beta_{2}} 
 \right|_{\beta_{1} = \beta_{2} = 1} 
= \frac{1}{N^{\frac{7}{4}}} 
\frac{\partial W(1,1;\alpha)}{\partial {\bar \alpha}}\ ,
 \label{one-loopcs-a2_CP2} \\
\frac{1}{N^{2}} \langle S \rangle
&=& \frac{1}{4} \left\langle \frac{1}{N} 
\tr (F_{\mu \nu})^2 \right\rangle + \alpha \langle M \rangle
= 2 \left( 1 - \frac{1}{N^{2}}\right) 
+ \frac{{\bar \alpha}}{4 N^{2}} \frac{\partial W(1,1;\alpha)}{
\partial {\bar \alpha}} \ . \nonumber \\
\label{one-loopacto2_CP2} 
%
\end{eqnarray}
Using the one-loop result
\begin{equation}
W(1,1;\alpha)_{\rm 1-loop} =
-\frac{1}{6} N^2 \alpha^4 (n-1)
+3 k^2 \sum_{s=1}^{m} (s+1)^3 
\log \left[ N \alpha^2 s(s+2) \right] - \log \{ {\rm vol}(H) {\cal N} \} \ ,
\end{equation}
which follows from (\ref{multi_cl_action_CP2}) and 
(\ref{coin_ol_CP2}),
we can reproduce (\ref{o2_CP2}) and (\ref{o1_CP2}).

\subsection{All order results from one-loop calculation}
 \label{allordercp2}

 Taking advantage of the one-loop saturation of
 the effective action mentioned at the end of appendix \ref{der_crit},
 we can obtain all order results for various quantities
 in the large-$N$ limit by simply shifting the center of expansion
 in the one-loop calculation \cite{0403242}.
 Similar calculations have been done previously 
 in the 3d YMCS model \cite{0410263}. 
 
 Since the free energy and the effective action
 are related to each other by the Legendre transformation, we can
 obtain the free energy by evaluating the effective action at its 
 extremum.  We consider the expansion around a configuration
 $A_{\mu} = \beta \, T_{\mu}^{(m,0)} \otimes {\bf 1}_{k}$.
 The value of ${\bar \beta}$ that gives the local minimum of the effective
 action can be obtained by solving 
   \begin{eqnarray}
    \frac{\partial \Gamma_{\rm 1-loop}}{\partial {\bar \beta}} 
 = N^{2} \left\{ \frac{2}{k}  ({\bar \beta}^{3}
    - {\bar \alpha} {\bar \beta}^{2}) 
    + \frac{6}{{\bar \beta}} \right\} =0  \ .
    \label{saddle-eq}
   \end{eqnarray}
%
 The solution exists for
 ${\bar \alpha} > {\bar \alpha}^{(k {\rm CP}^2)}_{\rm cr}
 = \frac{4}{\sqrt{3}} k^{\frac{1}{4}}$, and it is given explicitly as
  \begin{eqnarray}
   {\bar \beta} = f(\bar{\alpha}) \equiv
  \frac{\bar \alpha}{4} \left( 1 + \sqrt{1+ \delta} + \sqrt{2 - \delta
   + \frac{2}{\sqrt{1 + \delta}} }  \right), \label{beta-crit} 
  \end{eqnarray}
  where
  \begin{eqnarray}
   \delta = {\bar \alpha}^{-\frac{4}{3}} (96k)^{\frac{1}{3}} \left\{ \left( 1 
   + \sqrt{1 - \frac{256 k}{9 {\bar \alpha}^{4}}} \right)^{\frac{1}{3}}
   + \left(1 - \sqrt{1 - \frac{256 k}{9 {\bar \alpha}^{4}}} \right)^{\frac{1}{3}}
   \right\}.
  \end{eqnarray}
At large ${\bar \alpha}$, the solution (\ref{beta-crit}) can be expanded as
 \begin{eqnarray}
  f(\bar{\alpha})
= {\bar \alpha} \left( 1 - \sum_{j=1}^{\infty} c_{j} 
  {\bar \alpha}^{-4j} \right)
  = {\bar \alpha} 
 \left( 1 - \frac{3k}{{\bar \alpha}^{4}} - \frac{27k^{2}}{{\bar \alpha}^{8}}
  - \frac{405 k^{3}}{{\bar \alpha}^{12}} - \cdots \right).
  \label{beta-crit2}
 \end{eqnarray}
Plugging this solution into (\ref{eff-cp2-1loop}), 
we obtain the free energy to all orders as
 \begin{eqnarray}
  \frac{1}{N^{2}} W  
  \simeq - \frac{{\bar \alpha}^{4}}{6k} + 6 \log {\bar \alpha}
  + \log \frac{8N^{\frac{11}{2}}}{k^{3}}
  - \frac{9}{4}
  - \frac{9k}{{\bar \alpha}^{4}} - \frac{63k^{2}}{{\bar \alpha}^{8}}
  - \frac{1485 k^{3}}{2 {\bar \alpha}^{12}} - \cdots. \label{w-exact}
 \end{eqnarray}
 Using 
(\ref{one-loopf-sq2_CP2}), (\ref{one-loopcs-a2_CP2}) and 
(\ref{one-loopacto2_CP2}), we obtain the all order results
   \begin{eqnarray}
   \left\langle \frac{1}{N} \tr (F_{\mu\nu})^{2} \right\rangle &\simeq&
 \frac{2 \bar{\alpha}}{k} f(\bar{\alpha})^{3} + 8 = 
  \frac{2 {\bar \alpha}^{4}}{k} -10 
  - \frac{108k}{{\bar \alpha}^{4}} - \frac{1512 k^{2}}{{\bar \alpha}^{8}}
  - \frac{26730 k^{3}}{2 {\bar \alpha}^{12}} - \cdots. \nonumber \\ 
    \label{f-sq-exact} \\
   \frac{1}{N^{\frac{1}{4}}}  \langle M \rangle
  &\simeq& - \frac{2 f(\bar{\alpha})^{3}}{3k}
   = - \frac{2 {\bar \alpha}^{3}}{3k} + \frac{6}{{\bar \alpha}} 
   + \frac{36 k}{{\bar \alpha}^{5}}
   + \frac{504 k^{2}}{{\bar \alpha}^{9}} 
   + \frac{8910 k^{3}}{{\bar \alpha}^{13}} + \cdots \ . 
   \label{cs-a-exact} \\
  \frac{1}{N^{2}} \langle S \rangle &\simeq& 
 - \frac{1}{6k} \bar{\alpha} f(\bar{\alpha})^{3} + 2 
 = - \frac{{\bar \alpha}^{4}}{6k} + \frac{7}{2}
  + \frac{9k}{{\bar \alpha}^{4}} + \frac{126 k^{2}}{{\bar \alpha}^{8}}
  + \frac{4455 k^{3}}{2 {\bar \alpha}^{12}} + \cdots \ .
    \label{acto-exact}
 \end{eqnarray}
 
 Similarly to the case of the 3d YMCS model \cite{0410263}, 
 we can also calculate various observables directly
 to all orders in perturbation theory in the large-$N$ limit.
 For instance, the observable 
 $\frac{1}{\sqrt{N}} \Bigl\langle \frac{1}{N}
\tr (A_\mu)^{2} \Bigr\rangle$
 is obtained at one loop as (\ref{tr_A^2_largeN_CP2}),
 whose first and second terms correspond to 
 the classical and one-loop contributions, respectively.
 As we see from (\ref{tr_A^2_CP2}), however,
 the one-loop contribution comes from the tadpole
 diagram, which is not one-particle irreducible.
 Therefore, by replacing ${\bar \alpha}$ by $f({\bar \alpha})$ 
 in the classical contribution, 
 we obtain the all order result as
    \begin{eqnarray}
  \frac{1}{\sqrt{N}} \left\langle \frac{1}{N} \tr (A_\mu)^{2} \right\rangle 
  &\simeq& \frac{2 f(\bar{\alpha})^{2}}{3k} 
   = \frac{2{\bar \alpha}^{2}}{3k} - \frac{4}{{\bar \alpha}^{2}}
   - \frac{30k}{{\bar \alpha}^{6}}
   - \frac{432k^{2}}{{\bar \alpha}^{10}} 
   - \frac{7722 k^{3}}{{\bar \alpha}^{14}} - \cdots \ .
   \label{a-sq-exact}
 \end{eqnarray}
%
 
 \section{Perturbative calculations for the fuzzy ${\rm S}^2$}
\label{s2oneloop}
In this section we perform the perturbative analysis 
for the $k$ coincident ${\rm S}^2$ solution
  \begin{eqnarray}
  X_{\mu} = A^{ (k \, {\rm S}^{2})}_{\mu} \equiv
\left\{ \begin{array}{ll} \alpha \,  L^{(n)}_{\mu} \otimes {\bf 1}_k 
 & \textrm{~for $\mu=1,2,3$} \ , \\ 0 & \textrm{~otherwise}  \ ,
  \end{array} \right. 
\label{k-multi-fuzzyS2}
  \end{eqnarray}
which generalizes (\ref{cp2s2sol}).
The total size of the matrix is now given by $N = n k$. 
Let us calculate the free energy as
a perturbative expansion
$W = \sum_{j=0}^{\infty} W_{j}$,
where $W_{j} = {\rm O}(\alpha^{4(1-j)})$ comes from the $j$-loop contribution.
From (\ref{W1-def})
the free energy is obtained at the one-loop level as
\begin{eqnarray}
W_{0} + W_{1} &=& -\frac{\alpha^4 N^2}{24}(n^2-1)
+ 3k^2 \sum_{l=1}^{n-1} 
(2l+1)\log[N\alpha^2 l(l+1)] - \log \{ {\rm vol}(H) {\cal N} \} \n \\
&\simeq& N^2 \left( -\frac{\tilde\alpha^4}{24k^2}
+ 6 \log{\tilde\alpha}
+ \log \frac{N^7}{k^6} 
- \frac{15}{4}
\right) \ ,
\label{OL-effact_S2}
\end{eqnarray}
where the rescaled parameter $\tilde \alpha$ is defined by
(\ref{cp2s2rescaled}).

Similarly, the effective action for a one-parameter family of
configurations
  \begin{eqnarray}
A_{\mu} = 
\left\{ \begin{array}{ll} \beta \,  L^{(n)}_{\mu} \otimes {\bf 1}_k
 & \textrm{~for $\mu=1,2,3$} \ , \\ 0 & \textrm{~otherwise} 
  \end{array} \right. 
\label{k-multi-fuzzyS2-rescaled}
  \end{eqnarray}
can be obtained at the one-loop level as
 \begin{eqnarray}
\Gamma_{\rm 1-loop} (\tilde \beta)
\simeq N^2 \left(
\frac{{\tilde \beta}^{4}}{8k^{2}} 
   - \frac{{\tilde \alpha} {\tilde \beta}^{3}}{6k^{2}} 
   + 6 \log {\tilde \beta}
   + \log \frac{N^7}{k^6}
   - \frac{15}{4} \right) \ , 
\label{Gam-S2}
 \end{eqnarray}
 where $\tilde \beta = \beta N^{\frac{1}{2}} $.
The effective action has a local minimum
 \begin{eqnarray}
   {\tilde \beta} &=& g(\tilde{\alpha}) \equiv 
\frac{\tilde \alpha}{4} \left( 1 + \sqrt{1+ \varepsilon} 
  + \sqrt{2 - \varepsilon
   + \frac{2}{\sqrt{1 + \varepsilon}} }  \right) \ , \nonumber \\
   \varepsilon &=& {\tilde \alpha}^{-\frac{4}{3}} 
  (384 k^{2})^{\frac{1}{3}} 
  \left\{ \left( 1 
   + \sqrt{1 - \frac{1024 k^{2}}{9 {\tilde \alpha}^{4}}} \right)^{\frac{1}{3}}
   + \left(1 - \sqrt{1 - \frac{1024 k^{2}}{9 {\tilde \alpha}^{4}}} 
      \right)^{\frac{1}{3}} \right\} \ ,
\label{local-min-s2}
  \end{eqnarray}
if $\tilde{\alpha}$ is larger than the critical point
\begin{eqnarray}
   {\tilde \alpha}_{\rm cr}^{(k \, {\rm S}^{2})} = \sqrt{\frac{32k}{3}} 
   = 3.2659863 \cdots \times \sqrt{k} \ .
\label{anal-crit_S2}
\end{eqnarray}
Since the effective action is saturated at the one-loop level 
in the large-$N$ limit \cite{0303120},
the critical point obtained above should be correct to all orders
in perturbation theory.

The propagators for $\tilde A_\mu$ and the ghosts are 
exactly the same as (C.1) and (C.2) in ref.\ \cite{0401038}.
The tadpole is given by 
\begin{equation}
\langle \tilde A_\mu \rangle_{\rm 1-loop} = 
- \frac{12 k^2}{N^2 \alpha^3} X_\mu \ . 
\end{equation}
The one-loop results for various observables are obtained as
\begin{eqnarray}
 \frac{1}{N^{\frac{1}{2}}} \langle M \rangle_{\rm 1-loop} 
  &=& -\frac{\alpha^3 (n^2 -1)}{6N^{\frac{1}{2}}}
+ \frac{6}{\alpha N^{\frac{1}{2}}}\left(1-\frac{1}{n^2} \right) \n \\
&\simeq& -\frac{{\tilde \alpha}^3}{6k^{2}}
+ \frac{6}{{\tilde \alpha}} \ , \\
\left\langle \frac{1}{N}\tr (F_{\mu\nu})^2 
 \right\rangle_{\rm 1-loop} &=& 
 8 \left( 1-\frac{1}{N^2} \right)
-3\alpha \langle M \rangle_{\rm 1-loop} \n \\
&=&\frac{1}{2} \alpha^4 (n^2-1) - 10 +\frac{1}{N^2}
\Bigl\{8 (3k^2-1) -6k^2 \Bigr\} \n \\
&\simeq&\frac{\tilde \alpha^4}{2k^2} - 10 \ , \\
\frac{1}{N^2} \langle S \rangle_{\rm 1-loop} 
&=&\frac{1}{4} \left\langle \frac{1}{N}\tr (F_{\mu\nu})^2  
\right\rangle_{\rm 1-loop} 
+ \alpha \langle M \rangle_{\rm 1-loop} \n \\
&=&-\frac{\alpha^4}{24} (n^2 -1) + \frac{7}{2}
+\frac{1}{2N^2}\Bigl\{ - 4(k^2+1) + k^2 \Bigr\} \n \\
&\simeq& 
-\frac{\tilde\alpha^4}{24k^2} + \frac{7}{2} \ ,  \\
\frac{1}{N} \left\langle \frac{1}{N} \tr (A_\mu)^2
\right \rangle_{\rm 1-loop}
&=&
\alpha^2 \left\{ \frac{1}{4N}(n^2-1)
-\frac{6k^2 (n^2-1)}{N^3 \alpha^4} 
+\frac{8}{N n^2 \alpha^4} 
\sum_{l=1}^{n-1} \frac{2l+1}{l(l+1)} \right\} \n \\
&\simeq& \tilde\alpha^2 \left(
\frac{1}{4k^2} - \frac{6}{\tilde\alpha^4} + 
\frac{16}{n^2 \tilde\alpha^4} \log n \right) \n \\
&\simeq& \tilde\alpha^2 \left(
\frac{1}{4k^2} - \frac{6}{\tilde\alpha^4} \right) \ .
\end{eqnarray}

Exploiting the one-loop saturation of the effective action
in the large-$N$ limit,
we can calculate various quantities to all orders.
The free energy is obtained as
\begin{eqnarray}
   \frac{1}{N^{2}}  W
 &\simeq& - \frac{{\tilde \alpha}^{4}}{24k^{2}} + 6 \log {\tilde \alpha}
  + \log \frac{N^7}{k^6}
  - \frac{15}{4}
  - \frac{36k^{2}}{{\tilde \alpha}^{4}} 
  - \frac{1008k^{4}}{{\tilde \alpha}^{8}}
  - \frac{47520 k^{6}}{{\tilde \alpha}^{12}}
  - \cdots 
\nonumber \\ \label{w-exact-s2}
  \end{eqnarray}
by plugging (\ref{local-min-s2}) into (\ref{Gam-S2}).
Various observables are calculated as
  \begin{eqnarray}
    \frac{1}{N^{2}} \langle S \rangle &\simeq& 
  - \frac{1}{24 k^2} \tilde{\alpha} \, g(\tilde{\alpha})^3 + 2 = 
  - \frac{{\tilde \alpha}^{4}}{24k^{2}} + \frac{7}{2}
  + \frac{36k^{2}}{{\tilde \alpha}^{4}} 
  + \frac{2016 k^{4}}{{\tilde \alpha}^{8}}
  + \frac{142560 k^{6}}{{\tilde \alpha}^{12}} + \cdots \ , 
\nonumber \\ \label{acto-exact-s2}\\
    \frac{1}{N} 
  \left\langle \frac{1}{N} \tr (A_\mu)^{2} \right\rangle &\simeq&
  \frac{1}{4 k^2} g(\tilde{\alpha})^2 
  = \frac{{\tilde \alpha}^{2}}{4k^{2}} - \frac{6}{{\tilde \alpha}^{2}}
  - \frac{180k^{2}}{{\tilde \alpha}^{2}} 
  - \frac{10368 k^{4}}{{\tilde \alpha}^{6}}
  - \frac{741312k^{6}}{{\tilde \alpha}^{10}} - \cdots \ ,
    \label{a-sq-exact-s2} \\
        \frac{1}{N^{\frac{1}{2}}} \langle M \rangle &\simeq& 
   - \frac{1}{6 k^2} g(\tilde{\alpha})^3 
  = - \frac{{\tilde \alpha}^{3}}{6k^{2}} + \frac{6}{{\tilde \alpha}}
  + \frac{144k^{2}}{{\tilde \alpha}^{5}} 
  + \frac{8064 k^{4}}{{\tilde \alpha}^{9}}
  + \frac{570240 k^{6}}{{\tilde \alpha}^{13}} + \cdots \ , 
   \label{cs-a-exact-s2} \\
    \left\langle \frac{1}{N} \tr (F_{\mu\nu})^{2} \right\rangle &\simeq&
    \frac{1}{2 k^2} \tilde{\alpha} \, g(\tilde{\alpha})^3 + 8
  =  \frac{{\tilde \alpha}^{4}}{2k^{2}} - 10
  - \frac{432k^{2}}{{\tilde \alpha}^{4}} 
  - \frac{24192 k^{4}}{{\tilde \alpha}^{8}}
  - \frac{1710720k^{6}}{{\tilde \alpha}^{12}} 
  - \cdots \ .
\nonumber \\ \label{f-sq-exact-s2}
 \end{eqnarray}

\end{document}